\documentclass[11pt,preprintnumbers,titlepage,nofootinbib]{revtex4-2}%
\usepackage[latin9]{inputenc}
\usepackage{amsmath}
\usepackage{amssymb}
\usepackage{graphicx}
\usepackage{wasysym}
\usepackage[unicode=true,  bookmarks=false,  breaklinks=false,pdfborder={0 0
1},backref=section,colorlinks=false]{hyperref}
\usepackage{array}
\usepackage{multirow}
\usepackage{wasysym}
\usepackage{wasysym}
\usepackage{amsfonts}
\usepackage{subfigure}
\usepackage{cleveref}
\usepackage{listings}
\usepackage{xcolor}
\usepackage{array}
\usepackage{url}
\usepackage{color}%
\hypersetup{
colorlinks,linkcolor=blue,citecolor=blue,urlcolor=blue}
\makeatletter
\newcommand{\lyxdot}{.}
\providecommand{\tabularnewline}{\\}
\@ifundefined{textcolor}{}{
\definecolor{BLACK}{gray}{0}
\definecolor{WHITE}{gray}{1}
\definecolor{RED}{rgb}{1,0,0}
\definecolor{GREEN}{rgb}{0,1,0}
\definecolor{BLUE}{rgb}{0,0,1}
\definecolor{CYAN}{cmyk}{1,0,0,0}
\definecolor{MAGENTA}{cmyk}{0,1,0,0}
\definecolor{YELLOW}{cmyk}{0,0,1,0}
}

\makeatother
\begin{document}
\preprint{CTP-SCU/2023007}
\title{Gravitational Lensing by Born-Infeld Naked Singularities}
\author{Yiqian Chen$^{a}$}
\email{yqchen@stu.scu.edu.cn}
\author{Peng Wang$^{a}$}
\email{pengw@scu.edu.cn}
\author{Houwen Wu$^{a,b}$}
\email{hw598@damtp.cam.ac.uk}
\author{Haitang Yang$^{a}$}
\email{hyanga@scu.edu.cn}
\affiliation{$^{a}$Center for Theoretical Physics, College of Physics, Sichuan University,
Chengdu, 610064, China}
\affiliation{$^{b}$Department of Applied Mathematics and Theoretical Physics, University of
Cambridge, Wilberforce Road, Cambridge, CB3 0WA, UK}

\begin{abstract}
We examine the gravitational lensing phenomenon caused by photon spheres in
the Born-Infeld naked singularity spacetime, where gravity is coupled with
Born-Infeld electrodynamics. Specifically, our focus lies on relativistic
images originating from a point-like light source generated by strong
gravitational lensing near photon spheres, as well as images of a luminous
celestial sphere. It shows that Born-Infeld naked singularities consistently
exhibit one or two photon spheres, which project onto one or two critical
curves on the image plane. Interestingly, we discover that the nonlinearity
nature of the Born-Infeld electrodynamics enables photons to traverse the
singularity, leading to the emergence of new relativistic images within the
innermost critical curve. Furthermore, the presence of two photon spheres
doubles the number of relativistic images compared to the scenario with only a
single photon sphere. Additionally, the transparency inherent to Born-Infeld
naked singularities results in the absence of a central shadow in the images
of celestial spheres.

\end{abstract}
\maketitle
\tableofcontents

{}

\section{Introduction}

\label{sec:Introduction}

Gravitational lensing, the phenomenon of light bending in curved space, is a
captivating and fundamental effect predicted by general relativity
\cite{Dyson:1920cwa,Einstein:1936llh,Eddington:1987tk}. Due to its pivotal
role in astrophysics and cosmology, extensive research has been conducted on
gravitational lensing in the past decades. It has contributed significantly to
addressing crucial topics such as the distribution of structures
\cite{Mellier:1998pk,Bartelmann:1999yn,Heymans:2013fya}, dark matter
\cite{Kaiser:1992ps,Clowe:2006eq,Atamurotov:2021hoq}, dark energy
\cite{Biesiada:2006zf,Cao:2015qja,DES:2020ahh,DES:2021wwk}, quasars
\cite{SDSS:2000jpb,Peng:2006ew,Oguri:2010ns,Yue:2021nwt}, and gravitational
waves \cite{Seljak:2003pn,Diego:2021fyd,Finke:2021znb}. In an idealized lens
model involving a distant source in a Schwarzschild black hole, the slight
deflection of light in a weak gravitational field gives rise to the
observation of a primary and a secondary image. Moreover, strong gravitational
lensing near the photon sphere generates an infinite sequence of higher-order
images, known as relativistic images, on both sides of the optic axis
\cite{Virbhadra:1999nm}. Remarkably, relativistic images are minimally
affected by the characteristics of the astronomical source, making them
valuable for investigating the nature of the black hole spacetime.

Recently, the remarkable achievement of high angular resolution by the Event
Horizon Telescope collaboration
\cite{Akiyama:2019cqa,Akiyama:2019brx,Akiyama:2019sww,Akiyama:2019bqs,Akiyama:2019fyp,Akiyama:2019eap,Akiyama:2021qum,Akiyama:2021tfw,EventHorizonTelescope:2022xnr,EventHorizonTelescope:2022vjs,EventHorizonTelescope:2022wok,EventHorizonTelescope:2022exc,EventHorizonTelescope:2022urf,EventHorizonTelescope:2022xqj}%
, has facilitated the study of gravitational lensing in the strong gravity
regime, reigniting interest in the shadow of black hole images and the
associated phenomenon of strong gravitational lensing
\cite{Falcke:1999pj,Claudel:2000yi,Eiroa:2002mk,Virbhadra:2008ws,Yumoto:2012kz,Wei:2013kza,Zakharov:2014lqa,Atamurotov:2015xfa,Cunha:2016wzk,Dastan:2016bfy,Amir:2017slq,Wang:2017hjl,Ovgun:2018tua,Perlick:2018iye,Kumar:2019pjp,Zhu:2019ura,Ma:2019ybz,Mishra:2019trb,Zeng:2020dco,Zeng:2020vsj,Saurabh:2020zqg,Roy:2020dyy,Li:2020drn,Kumar:2020hgm,Zhang:2020xub,Guerrero:2022qkh,Virbhadra:2022iiy}%
. It has been demonstrated that strong gravitational lensing exhibits a close
connection to bound photon orbits, which give rise to photon spheres in
spherically symmetric black holes. Intriguingly, certain horizonless
ultra-compact objects have been discovered to harbor photon spheres,
effectively mimicking black holes in numerous observational simulations
\cite{Schmidt:2008hc,Guzik:2009cm,Liao:2015uzb,Goulart:2017iko,Nascimento:2020ime,Qin:2020xzu,Junior:2021svb,Islam:2021ful,Tsukamoto:2021caq,Olmo:2021piq}%
.  Among these objects, naked singularities have garnered significant
attention. Although the cosmic censorship conjecture forbids the formation of
naked singularities, it is possible for these entities to arise through the
gravitational collapse of massive objects under specific initial conditions
\cite{Shapiro:1991zza,Joshi:1993zg,Harada:1998cq,Joshi:2001xi,Goswami:2006ph,Banerjee:2017njk,Bhattacharya:2017chr}%
. Given that the presence of photon spheres allows naked singularities to
emulate the optical appearance of black holes, the gravitational lensing
phenomena associated with naked singularities have been extensively
investigated
\cite{Virbhadra:2002ju,Virbhadra:2007kw,Gyulchev:2008ff,Sahu:2012er,Banerjee:2018clz,Shaikh:2019itn,Paul:2020ufc,Tsukamoto:2021fsz}%
.

In the context of Reissner-Nordstr\"{o}m (RN) naked singularities
characterized by a mass $M$ and charge $Q$,\ it is noteworthy that a photon
sphere exists only if $1<Q/M\leq\sqrt{9/8}$, whereas no photon sphere is
present when  $Q/M>\sqrt{9/8}$. The phenomenon of strong gravitational lensing
by the photon sphere has been investigated within the spacetime of RN naked
singularities \cite{Shaikh:2019itn,Tsukamoto:2021fsz}. Analogous to the case
of black holes, two sets of relativistic images can be observed beyond the
critical curve, which arises from photons originating from the photon sphere.
Remarkably, two additional sets of brighter relativistic images have been
identified within the critical curve due to the absence of an event horizon
and the existence of a potential barrier near the anti-photon sphere.
Furthermore, as demonstrated below, when a celestial sphere illuminates an RN
naked singularity, the absence of a shadow at the center of the image is
observed since light rays entering the photon sphere are reflected by the
potential barrier at the singularity. These distinctive observational
characteristics can serve as means to differentiate between RN singularities
and RN black holes.

The Born-Infeld electrodynamics was initially proposed to regulate the
divergences arising from the electrostatic self-energy of point charges,
achieved through the introduction of an electric field cutoff
\cite{Born:1934gh}. Subsequently, it was recognized that Born-Infeld
electrodynamics can emerge from the low-energy limit of string theory, which
describes the dynamics of D-branes at low energies. Coupling the Born-Infeld
electrodynamics field to gravity, the Born-Infeld black hole solution was
obtained in \cite{Dey:2004yt,Cai:2004eh}. Since then, a multitude of
properties pertaining to Born-Infeld black holes have been extensively
examined
\cite{Fernando:2003tz,Banerjee:2010da,Zou:2013owa,Hendi:2015hoa,Zeng:2016sei,Li:2016nll,Tao:2017fsy,Dehyadegari:2017hvd,Wang:2018xdz,Liang:2019dni,Gan:2019jac,Wang:2019kxp,Wang:2020ohb}%
. More recently, it has been reported that Born-Infeld naked singularity
solutions can exhibit two photon spheres within specific parameter ranges
\cite{Guo:2022ghl}.

The primary objective of this paper is to investigate gravitational lensing
phenomena exhibited by Born-Infeld naked singularities. Remarkably, our
findings unveil the ability of photons to traverse these singularities, which,
in conjunction with the presence of double photon spheres, gives rise to
distinct observational signatures. The subsequent sections of this paper are
structured as follows: In Section \ref{sec:BINS}, we provide a concise
overview of the Born-Infeld naked singularity solution and present their
domain of existence. Section \ref{sec:PT} focuses on the analysis of photon
trajectories in an effective geometry and explores their behavior in proximity
to the singularity. The discussion then proceeds to examine relativistic
images of a distant light source in Section
\ref{sec:Strong Gravitational Lensing}, followed by the analysis of images
produced by a luminous celestial sphere in Section
\ref{sec:Observational Images}. Finally, Section \ref{sec:CONCLUSIONS}
presents our conclusions. We adopt the convention $16\pi G=c=1$ throughout the paper.

\section{Born-Infeld Naked Singularity}

\label{sec:BINS}

We consider a $\left(  3+1\right)  $ dimensional gravity model coupled to a
Born-Infeld electromagnetic field $A_{\mu}$. The action $\mathcal{S}$ is given
by
\begin{equation}
\mathcal{S}=\int d^{4}x\sqrt{-g}\left[  R+4\mathcal{L}\left(  s,p\right)
\right]  , \label{eq:NLEDAction}%
\end{equation}
where
\begin{equation}
\mathcal{L}\left(  s,p\right)  =\frac{1}{a}\left(  1-\sqrt{1-2as-a^{2}p^{2}%
}\right)  \text{.}%
\end{equation}
Here, $s$ and $p$ are two independent nontrivial scalars constructed from the
field strength tensor $F_{\mu\nu}=\partial_{\mu}A_{\nu}-\partial_{\nu}A_{\mu}$
and none of its derivatives, i.e.,
\begin{equation}
s=-\frac{1}{4}F^{\mu\nu}F_{\mu\nu}\text{ and }p=-\frac{1}{8}\epsilon^{\mu
\nu\rho\sigma}F_{\mu\nu}F_{\rho\sigma}\text{,}%
\end{equation}
where $\epsilon^{\mu\nu\rho\sigma}\equiv-\left[  \mu\text{ }\nu\text{ }%
\rho\text{ }\sigma\right]  /\sqrt{-g}$ is a totally antisymmetric Lorentz
tensor, and $\left[  \mu\text{ }\nu\text{ }\rho\text{ }\sigma\right]  $
denotes the permutation symbol. The coupling parameter $a$ is related to the
string tension $\alpha^{\prime}$ as $a=\left(  2\pi\alpha^{\prime}\right)
^{2}$. In the limit $a\rightarrow0$, the Born-Infeld Lagrangian $\mathcal{L}%
\left(  s,p\right)  $ reduces to the Lagrangian of the Maxwell field. The
equations of motion can be obtained by varying the action $\left(
\ref{eq:NLEDAction}\right)  $ with respect to $g_{\mu\nu}$ and $A_{\mu}$,
yielding
\begin{align}
R_{\mu\nu}-\frac{1}{2}Rg_{\mu\nu}  &  =\frac{T_{\mu\nu}}{2}\text{,}\nonumber\\
\nabla_{\mu}\left[  \frac{\partial\mathcal{L}\left(  s,p\right)  }{\partial
s}F^{\mu\nu}+\frac{1}{2}\frac{\partial\mathcal{L}\left(  s,p\right)
}{\partial p}\epsilon^{\mu\nu\rho\sigma}F_{\rho\sigma}\right]   &  =0\text{,}
\label{eq:NLEDEOM}%
\end{align}
where $T_{\mu\nu}$ is the energy-momentum tensor,
\begin{equation}
T_{\mu\nu}=4g_{\mu\nu}\left[  \mathcal{L}\left(  s,p\right)  -p\frac
{\partial\mathcal{L}\left(  s,p\right)  }{\partial p}\right]  +\frac
{\partial\mathcal{L}\left(  s,p\right)  }{\partial s}F_{\mu}^{\text{ }\rho
}F_{\nu\rho}\text{.}%
\end{equation}

The spherically symmetric ansatz yields a solution to the equations of motion
$\left(  \ref{eq:NLEDEOM}\right)  $ \cite{Dey:2004yt,Cai:2004eh,Guo:2022ghl}.
The metric is given by
\begin{align}
ds^{2} &  =g_{\mu\nu}dx^{\mu}dx^{\nu}=-f_{\text{BI}}\left(  r\right)
dt^{2}+\frac{dr^{2}}{f_{\text{BI}}\left(  r\right)  }+r^{2}\left(  d\theta
^{2}+\sin^{2}\theta d\varphi^{2}\right)  \text{,}\nonumber\\
A &  =A_{t}\left(  r\right)  dt-P\cos\theta d\varphi,\label{eq:NLEDBH}%
\end{align}
where
\begin{align}
f_{\text{BI}}\left(  r\right)   &  =1-\frac{2M}{r}-\frac{2\left(  Q^{2}%
+P^{2}\right)  }{3\sqrt{r^{4}+a\left(  Q^{2}+P^{2}\right)  }+3r^{2}}%
+\frac{4\left(  Q^{2}+P^{2}\right)  }{3r^{2}}\text{ }_{2}F_{1}\left(  \frac
{1}{4},\frac{1}{2},\frac{5}{4};-\frac{a\left(  Q^{2}+P^{2}\right)  }{r^{4}%
}\right)  ,\nonumber\\
A_{t}^{\prime}\left(  r\right)   &  =\frac{Q}{\sqrt{r^{4}+a\left(  Q^{2}%
+P^{2}\right)  }}.
\end{align}
The mass, electrical charge, and magnetic charge of the black hole are denoted
by $M$, $Q$ and $P$, respectively, and $_{2}F_{1}\left(  a,b,c;x\right)  $ is
the hypergeometric function. Moreover, the solution appears to have a
singularity at $r=0$. The nature of the singularity is investigated using the
Kretschmann scalar $\mathcal{K}=R^{\mu\nu\rho\sigma}R_{\mu\nu\rho\sigma}$. Our
calculation reveals that the origin is a physical singularity as
\begin{equation}
\mathcal{K=}\frac{16}{3\pi r^{6}}\left[  3M\sqrt{\pi}-2a^{-1/4}\left(
Q^{2}+P^{2}\right)  ^{2}\Gamma\left(  1/4\right)  \Gamma\left(  5/4\right)
\right]  ^{2}+\mathcal{O}\left(  r^{-5}\right)  .
\end{equation}
Thus, the solution $\left(  \ref{eq:NLEDBH}\right)  $ describes a naked
singularity at $r=0$ or a black hole if an event horizon exists.

\begin{figure}[t]
\centering\begin{minipage}[c]{0.45\linewidth}%
\subfigure{\includegraphics[scale=0.65]{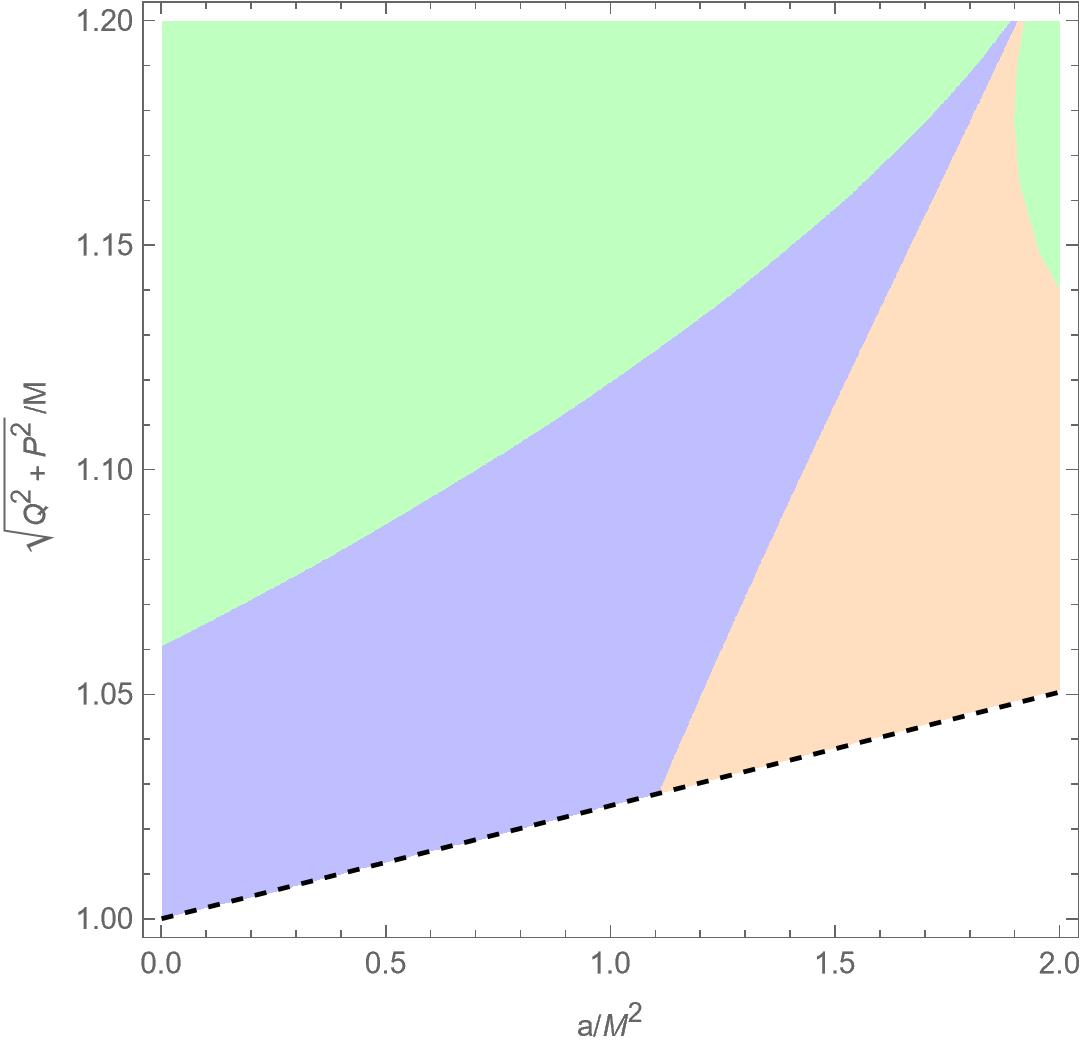}} %
\end{minipage}\hspace{20pt} \begin{minipage}[c]{0.45\linewidth}%
\subfigure{\includegraphics[scale=0.45]{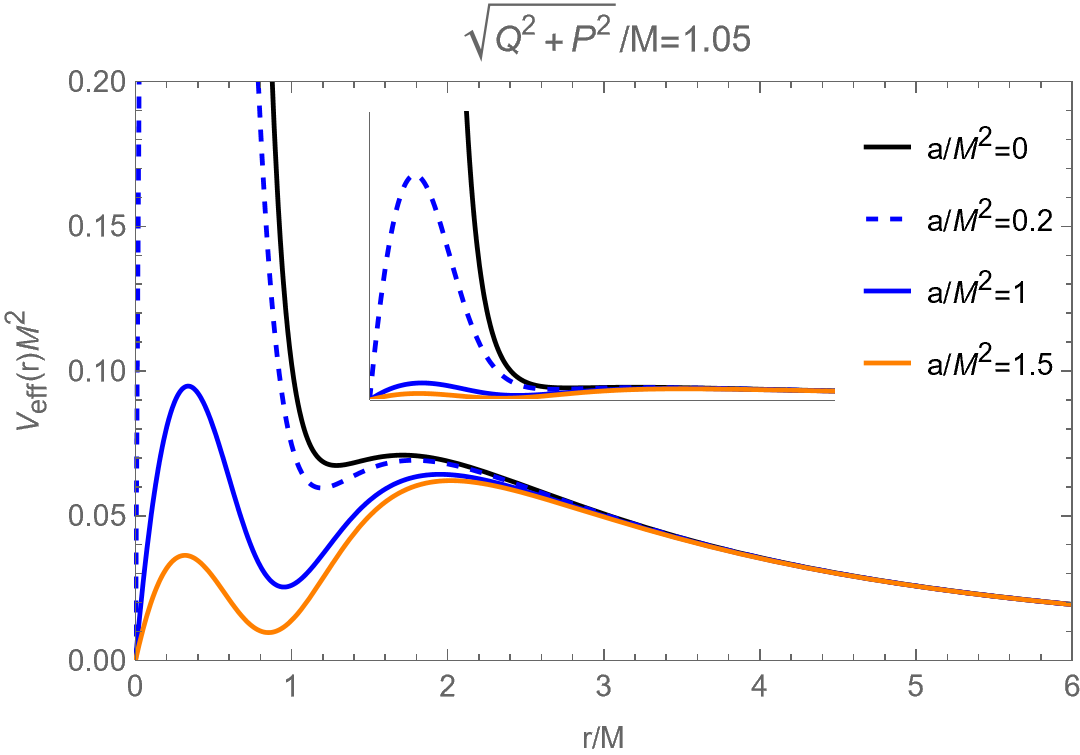}}\\
\subfigure{\includegraphics[scale=0.45]{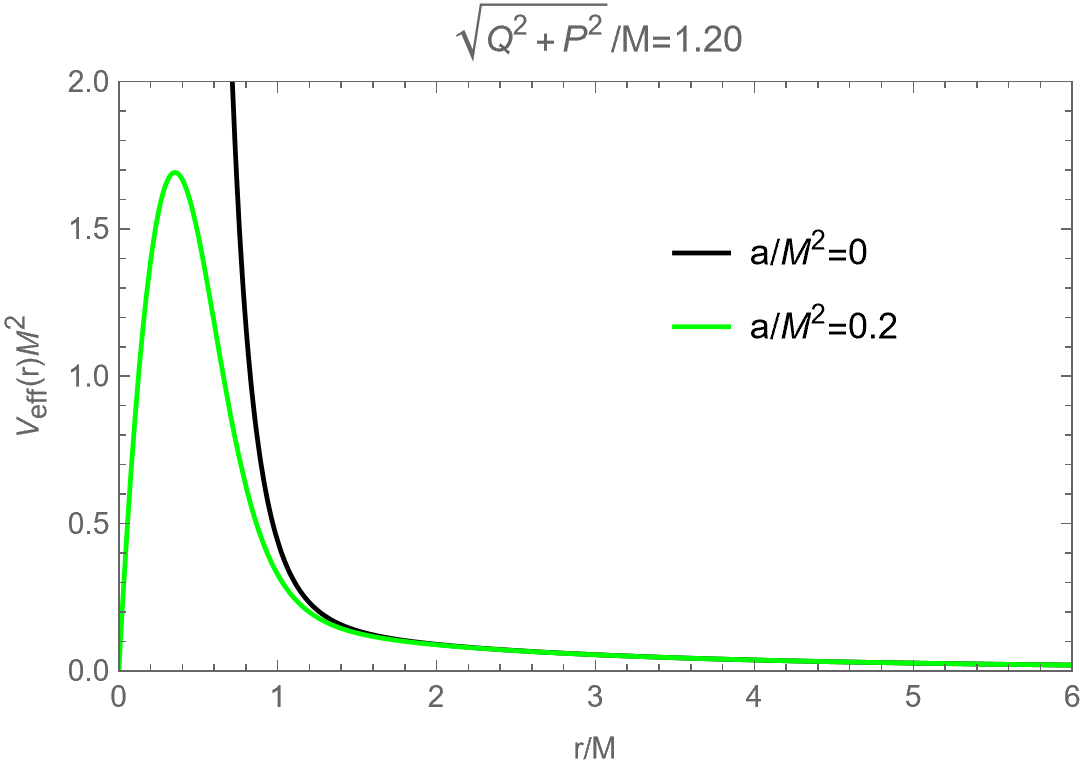}} %
\end{minipage}\caption{\textbf{Left}: The Born-Infeld metric $\left(
\ref{eq:NLEDBH}\right)  $ above the dashed black line describes a naked
singularity at $r=0$. Within the blue and orange regions, naked singularity
solutions have two photon spheres, and the effective potential at the inner
photon sphere is higher/lower than that at the outer one in the blue/orange
regions. The solutions in the green region have only one photon sphere.
\textbf{Right}: The upper and lower panels show the effective potential
$V_{\text{eff}}\left(  r\right)  $ of representative naked singularity
solutions with $Q=1.05$ and $1.20$, respectively. A photon sphere corresponds
to a local maximum of $V_{\text{eff}}\left(  r\right)  $.}%
\label{fig:BI}%
\end{figure}

To determine the separatrix between naked singularity and black hole
solutions, we investigate extremal black holes with horizon radius $r_{e}$ and
mass $M_{e}$. The conditions $f\left(  r_{e}\right)  =0=d\left(  rf\left(
r\right)  \right)  /dr|_{r=r_{e}}$ yield the expressions for $r_{e}$ and mass
$M_{e}$ as
\begin{align}
r_{e} &  =\frac{\sqrt{4\left(  Q^{2}+P^{2}\right)  -a}}{2},\nonumber\\
M_{e} &  =\frac{\sqrt{4\left(  Q^{2}+P^{2}\right)  -a}}{6}+\frac{8\left(
Q^{2}+P^{2}\right)  \,}{6\sqrt{4\left(  Q^{2}+P^{2}\right)  -a}}\text{ }%
_{2}F_{1}\left(  \frac{1}{4},\frac{1}{2};\frac{5}{4};-\frac{16a\left(
Q^{2}+P^{2}\right)  }{\left[  4\left(  Q^{2}+P^{2}\right)  -a\right]  ^{2}%
}\right)  .
\end{align}
It is evident that extremal black holes do not exist for $a>4\left(
Q^{2}+P^{2}\right)  $. So when $a<4\left(  Q^{2}+P^{2}\right)  $ and $M<M_{e}%
$, the spacetime is a naked singularity. However, when $a>4\left(  Q^{2}%
+P^{2}\right)  $, the spacetime can have at most one horizon. The presence of
the horizon can be determined by investigating $rf\left(  r\right)  $, which
vanishes at the horizon radius. In fact, one finds $d\left(  rf\left(
r\right)  \right)  /dr>0$ and $\lim\limits_{r\rightarrow0}rf\left(  r\right)
=4\left[  a\left(  Q^{2}+P^{2}\right)  \right]  ^{3/4}\Gamma\left(
1/4\right)  \Gamma\left(  5/4\right)  /\left(  3a\sqrt{\pi}\right)  -2M$,
indicating the appearance of a naked singularity when
\begin{equation}
M<\frac{2\Gamma\left(  1/4\right)  \Gamma\left(  5/4\right)  \left[  a\left(
Q^{2}+P^{2}\right)  \right]  ^{3/4}}{3a\sqrt{\pi}}.
\end{equation}
The left panel of FIG. \ref{fig:BI} shows the domain of existence for
Born-Infeld naked singularities in the $a/M^{2}$-$\sqrt{Q^{2}+P^{2}}/M$
parameter space, with the dashed black line denoting the separatrix between
the black hole and naked singularity solutions. Solutions above the dashed
black line in the colored regions represent Born-Infeld naked singularities.

\section{Photon Trajectories}

\label{sec:PT}

Nonlinear electrodynamics theories allow for self-interaction of the
electromagnetic field, leading to changes in the direction of photon
propagation and deviation from null geodesics. Propagation equations
describing photon trajectories can be obtained by analyzing the
electromagnetic field's discontinuity at the characteristic surface of wave
propagation. An effective metric is then introduced, in which photons travel
on null geodesics \cite{Novello:1999pg}. In one-parameter theories with the
Lagrangian as a function of $s$, a single effective geometry determines the
photon trajectories. However, in two-parameter theories with the Lagrangian as
a function of $s$ and $p$, two possible solutions exist, leading to
birefringence. The Born-Infeld theory, on the other hand, ensures the
uniqueness of the photon path via its equations of motion, and the effective
metric $\tilde{g}^{\mu\nu}$ is given by \cite{Novello:1999pg}
\begin{equation}
\tilde{g}^{\mu\nu}=\frac{\left(  1-2as\right)  g^{\mu\nu}+ag_{\rho\sigma
}F^{\mu\rho}F^{\sigma\nu}}{1-2as-a^{2}p^{2}}. \label{eq:geff}%
\end{equation}
Using the underlying Born-Infeld metric $\left(  \ref{eq:NLEDBH}\right)  $,
the effective metric takes the form
\begin{align}
d\tilde{s}^{2}  &  =\tilde{g}_{\mu\nu}dx^{\mu}dx^{\nu}=-f(r)dt^{2}%
+\frac{dr^{2}}{h(r)}+R(r)\left(  d\theta^{2}+\sin^{2}\theta d\varphi
^{2}\right) \nonumber\\
&  =\frac{(aP^{2}+r^{4})^{2}}{r^{2}\left[  a\left(  Q^{2}+P^{2}\right)
+r^{4}\right]  ^{3/2}}\left[  -f_{\text{BI}}(r)dt^{2}+\frac{dr^{2}%
}{f_{\text{BI}}(r)}+\frac{a\left(  Q^{2}+P^{2}\right)  +r^{4}}{r^{2}}\left(
d\theta^{2}+\sin^{2}\theta d\varphi^{2}\right)  \right]  , \label{eq:effM}%
\end{align}
where $\tilde{g}_{\mu\rho}\tilde{g}^{\rho\nu}=\delta_{\mu}^{\nu}$, and
\begin{align}
f(r)  &  =\frac{(aP^{2}+r^{4})^{2}}{r^{2}\left[  a\left(  Q^{2}+P^{2}\right)
+r^{4}\right]  ^{3/2}}f_{\text{BI}}(r),\nonumber\\
h(r)  &  =\frac{r^{2}\left[  a\left(  Q^{2}+P^{2}\right)  +r^{4}\right]
^{3/2}}{(aP^{2}+r^{4})^{2}}f_{\text{BI}}(r)\text{,}\\
R(r)  &  =\frac{(aP^{2}+r^{4})^{2}}{r^{4}\sqrt{a\left(  Q^{2}+P^{2}\right)
+r^{4}}}.\nonumber
\end{align}
Although the effective metric appears to lack electric-magnetic duality, this
symmetry is present when the metric is multiplied by a conformal factor
$(aP^{2}+r^{4})^{-2}$, which does not alter null geodesics. Therefore, the
electric-magnetic duality of photon trajectories is expected.

In the Hamiltonian canonical formalism, a photon with 4-momentum vector
$p^{\mu}=(\dot{t},\dot{r},\dot{\theta},\dot{\varphi})$, where dots stand for
derivative with respect to some affine parameter $\lambda$, has canonical
momentum $q_{\mu}=\tilde{g}_{\mu\nu}p^{\nu}$, which satisfies the null
condition $p^{\mu}q_{\mu}=0$. The null geodesic equations in the effective
metric $\left(  \ref{eq:effM}\right)  $ are separable and can be fully
characterized by three conserved quantities,
\begin{equation}
E=q_{\mu}\partial_{t}^{\mu}=-q_{t},\text{ }L_{z}=q_{\mu}\partial_{\varphi
}^{\mu}=q_{\varphi},\text{ }L^{2}=K^{\mu\nu}q_{\mu}q_{\nu}=q_{\theta}%
^{2}+L_{z}^{2}\csc^{2}\theta,\label{eq:ELK}%
\end{equation}
which denote the total energy, the angular momentum parallel to the axis of
symmetry, and the total angular momentum, respectively. Here, the tensor
$K^{\mu\nu}$ is an symmetric Killing tensor
\begin{equation}
K=R^{2}(r)\left(  d\theta\otimes d\theta+\sin^{2}\theta d\varphi\otimes
d\varphi\right)  .
\end{equation}
Note that $\tilde{\nabla}_{(\lambda}K_{\mu\nu)}=0$, where $\tilde{\nabla}$ is
the covariant derivative compatible with the effective metric. The canonical
4-momentum $q=q_{\mu}dx^{\mu}$ can be expressed in terms of $E$, $L_{z}$ and
$L$ as
\begin{align}
q &  =-Edt\pm_{r}\sqrt{\mathcal{R}(r)}dr\pm_{\theta}\sqrt{\Theta(\theta
)}d\theta+Ld\varphi,\nonumber\\
\mathcal{R}(r) &  =\frac{1}{h(r)}\left(  \frac{E^{2}}{f(r)}-\frac{L^{2}}%
{R(r)}\right)  \text{ and }\Theta(\theta)=L^{2}-L_{z}^{2}\csc^{2}\theta,
\end{align}
where the two choices of sign $\pm_{r}$ and $\pm_{\theta}$ depend on the
radial and polar directions of travel, respectively. Then, null geodesic
equations are given by $p^{\mu}=\tilde{g}^{\mu\nu}q_{\nu}$, i.e.,
\begin{equation}
\dot{t}=\frac{E}{f(r)},\text{ }\dot{r}=\pm_{r}L\sqrt{\frac{h(r)}{f(r)}\left[
b^{-2}-V_{\text{eff}}(r)\right]  },\text{ }\dot{\theta}=\pm_{\theta}%
\frac{\sqrt{L^{2}-L_{z}^{2}\csc^{2}\theta}}{R(r)},\text{ }\dot{\varphi}%
=\frac{L_{z}}{R(r)\sin^{2}(\theta)}\text{,}\label{eq:geo-eq}%
\end{equation}
where $b\equiv L/E$ is the impact parameter, and the effective potential of
photons in the effective metric is defined as
\begin{equation}
V_{\text{eff}}(r)=\frac{f(r)}{R(r)}.
\end{equation}

In the study of black holes, the near-singularity behavior of photons that
have entered the horizon is usually not considered. However, since naked
singularities lack horizons, it is crucial to investigate the behavior of
photons in the vicinity of the singularities. Eqn. $\left(  \ref{eq:geo-eq}%
\right)  $ gives the behavior of photons around the singularity at $r=0$,
yielding
\begin{equation}
\frac{dr}{dt}=\pm_{r}\frac{4\Gamma\left(  1/4\right)  \Gamma\left(
5/4\right)  \left(  Q^{2}+P^{2}\right)  ^{3/2}-6\sqrt{\pi}a^{1/4}M}{3\sqrt
{\pi}a^{1/4}r}+\mathcal{O}(r^{0}),
\end{equation}
which implies that a photon can pass through the singularity within a finite
coordinate time. Furthermore, we find that, near the singularity, solutions of
the null geodesic equations $\left(  \ref{eq:geo-eq}\right)  $ can be expanded
as
\begin{equation}
x^{\mu}=x_{0}^{\mu}+\sum\limits_{n=1}^{\infty}\sum\limits_{m=0}^{n-1}%
c_{nm}^{\mu}\lambda^{-n}\log^{m}\left\vert \lambda\right\vert \text{ for }%
\mu=t\text{, }r\text{, }\theta\text{ and }\varphi\text{,}\label{eq:expansion}%
\end{equation}
where $x_{0}^{\mu}$ are the constant of integration, and the coefficients
$c_{nm}^{\mu}$ are calculated recursively order by order. Particularly, the
leading coefficients are given by
\begin{align}
c_{20}^{t} &  =\pm_{r}\frac{3\sqrt{\pi}a^{5/4}\left(  Q^{2}+P^{2}\right)
}{8\Gamma\left(  1/4\right)  \Gamma\left(  5/4\right)  \left(  Q^{2}%
+P^{2}\right)  ^{3/2}E^{2}-12\sqrt{\pi}a^{1/4}E^{2}M},\text{ }c_{10}^{r}%
=\mp_{r}\frac{\sqrt{a\left(  Q^{2}+P^{2}\right)  }}{E},\nonumber\\
c_{30}^{\theta} &  =\mp_{\theta}\frac{\sqrt{a\left(  Q^{2}+P^{2}\right)
}\sqrt{L^{2}-L_{z}^{2}\csc^{2}x_{0}^{\theta}}}{3E^{4}},\text{ }c_{30}%
^{\varphi}=-\frac{\sqrt{a\left(  Q^{2}+P^{2}\right)  }L_{z}\csc^{2}%
x_{0}^{\theta}}{3E^{4}}.\label{eq:asy-solutions}%
\end{align}
As a light ray traverses the singularity, it splits into two branches, namely
the radially outgoing one associated with the upper sign of $\pm_{r}$ and
$\mp_{r}$, and the radially ingoing one with the lower sign. We adopt
$\lambda>0$ and $\lambda<0$ for the ingoing and outgoing branches,
respectively. It is worth emphasizing that\ the affine parameter approaches
$\pm\infty$ when the light ray approaches the singularity, leading to
$x_{0}^{r}=0$. At the singularity, the ingoing and outgoing branches are
connected by the conditions
\begin{equation}
t(-\infty)=t(\infty),\quad\theta(-\infty)=\pi-\theta(\infty),\quad
\varphi(-\infty)=\pi+\varphi(\infty).\label{eq:cc}%
\end{equation}

The effective potential $V_{\text{eff}}(r)$ determines the locations of
circular light rays, with unstable and stable light rays corresponding to
local maxima and minima, respectively. The unstable circular light rays form
photon spheres, which are critical for observing black holes. From eqn.
$\left(  \ref{eq:effM}\right)  $, it follows that $V_{\text{eff}}\left(
\infty\right)  =0=V_{\text{eff}}\left(  0\right)  $ when $a>0$, indicating the
existence of at least one photon sphere in the Born-Infeld naked singularity
spacetime. The left panel of FIG. \ref{fig:BI} illustrates the regions where
one or two photon spheres exist in the $a/M^{2}$-$\sqrt{Q^{2}+P^{2}}/M$
parameter space:

\begin{itemize}
\item The green region corresponds to a single photon sphere in the naked
singularity spacetime, as illustrated by the green line in the lower-right
panel of FIG. \ref{fig:BI} for $V_{\text{eff}}(r)$ with $\sqrt{Q^{2}+P^{2}%
}/M=1.20$ and $a/M^{2}=0.2$.

\item The blue regions correspond to naked singularities with two photon
spheres, as illustrated by the blue lines in the upper-right panel of FIG.
\ref{fig:BI} for $V_{\text{eff}}(r)$ with $\sqrt{Q^{2}+P^{2}}/M=1.05$ and
$a/M^{2}=0.2$ and $1$. The potential peak at the inner photon sphere is higher
than that at the outer one, indicating that both photon spheres play a role in
determining the optical appearances of luminous matters
\cite{Gan:2021xdl,Gan:2021pwu,Guo:2022muy,Chen:2022qrw}.

\item The orange regions correspond to naked singularities with two photon
spheres, as illustrated by the orange line in the upper-right panel of FIG.
\ref{fig:BI} for $V_{\text{eff}}(r)$ with $\sqrt{Q^{2}+P^{2}}/M=1.05$ and
$a/M^{2}=$ $1.5$. The potential peak at the inner photon sphere is lower than
that at the outer one. In this case, the inner photon sphere is invisible to
distant observers since light rays near it can not escape to infinity.
Nevertheless, the inner photon sphere is closely related to long-lived
quasinormal modes \cite{Guo:2021enm}, echo signals \cite{Guo:2022umh} and
superradiant instability \cite{Guo:2023ivz}.
\end{itemize}

The effective potential $V_{\text{eff}}(r)$ of RN naked singularities with
$a=0$ possesses a photon sphere provided that\ $1<\sqrt{Q^{2}+P^{2}}%
/M<\sqrt{9/8}$ (e.g., indicated by the black line in the upper-right panel of
FIG. \ref{fig:BI}), while no photon sphere exists when $\sqrt{Q^{2}+P^{2}%
}/M\geq\sqrt{9/8}$ (e.g., the black line in the lower-right panel of FIG.
\ref{fig:BI}) \cite{Pugliese:2010ps}. In addition, $V_{\text{eff}}(r)$
diverges at $r=0$ for RN naked singularities, preventing photons from reaching
the singularity. Interestingly, when the effects of nonlinear electrodynamics
are present, $V_{\text{eff}}(r)$ approaches zero instead of infinity as
$r\rightarrow0$, enabling photons with a sufficiently small impact parameter
to overcome the potential barrier and reach the singularity.

\section{Relativistic Images}

\label{sec:Strong Gravitational Lensing}

In this section, we explore the phenomenon of gravitational lensing caused by
Born-Infeld naked singularities in the context of the strong deflection limit.
Our analysis starts with determining the deflection angle, which allows us to
derive the angular positions of relativistic images. We employ an idealized
thin lens model that assumes a high degree of alignment among the source, lens
and observer. The lens equation, as presented in \cite{Virbhadra:1999nm}, is
expressed as
\begin{equation}
\beta=\vartheta-\frac{D_{LS}}{D_{OS}}\Delta\alpha,\label{eq:lens equation}%
\end{equation}
where $\beta$ represents the angular separation between the source and the
lens, $\vartheta$ denotes the angular separation between the lens and the
image, and $\Delta\alpha$ represents the offset of the deflection angle after
accounting for all the windings experienced by the photon. Here, the distances
$D_{OL}$, $D_{LS}$ and $D_{OS}$ correspond to the observer-lens, lens-source
and observer-source distances, respectively.

For the sake of simplicity, we confine our analysis to the equatorial plane,
taking advantage of the spherical symmetry. In the idealized model, the
deflection angle $\alpha(b)$ is described by the following expression from
\cite{Virbhadra:1999nm},
\begin{equation}
\alpha(b)=I(b)-\pi,\label{eq:deflection angle}%
\end{equation}
where $I(b)$ represents the change in $\varphi$, and $b$ denotes the impact
parameter related to $\vartheta$ through the equation $b=D_{OL}\vartheta$.
When a photon approaches a turning point at $r=r_{0}$ and then gets deflected
towards a distant observer, the integral $I(b)$ is given by
\begin{equation}
I(b)=2\int_{r_{0}}^{\infty}\frac{1}{\sqrt{h(r)R(r)\left[  R(r)/b^{2}%
f(r)-1\right]  }}dr.\label{eq:integral-reflect}%
\end{equation}
Alternatively, if the photon passes through the singularity at $r=0$, the
azimuthal angle $\varphi$ increases by $\pi$, resulting in the expression,
\begin{equation}
I(b)=2\int_{0}^{\infty}\frac{1}{\sqrt{h(r)R(r)\left[  R(r)/b^{2}f(r)-1\right]
}}dr+\pi.\label{eq:integral-pass}%
\end{equation}
In the strong deflection limit, the integral $I(b)$ diverges as the impact
parameter $b$ approaches the critical value $b_{c}$, which represents the
impact parameter for photon trajectories on the photon sphere at $r=r_{c}$. By
expanding $I(b)$ around $b=b_{c}$ (or, equivalently $r_{0}=r_{c}$), we can
obtain $\alpha(b)$ in the strong deflection limit.

\subsection{Single Photon Sphere}

\label{subsec:Single-peak potential}

\begin{figure}[ptb]
\includegraphics[width=0.88\textwidth]{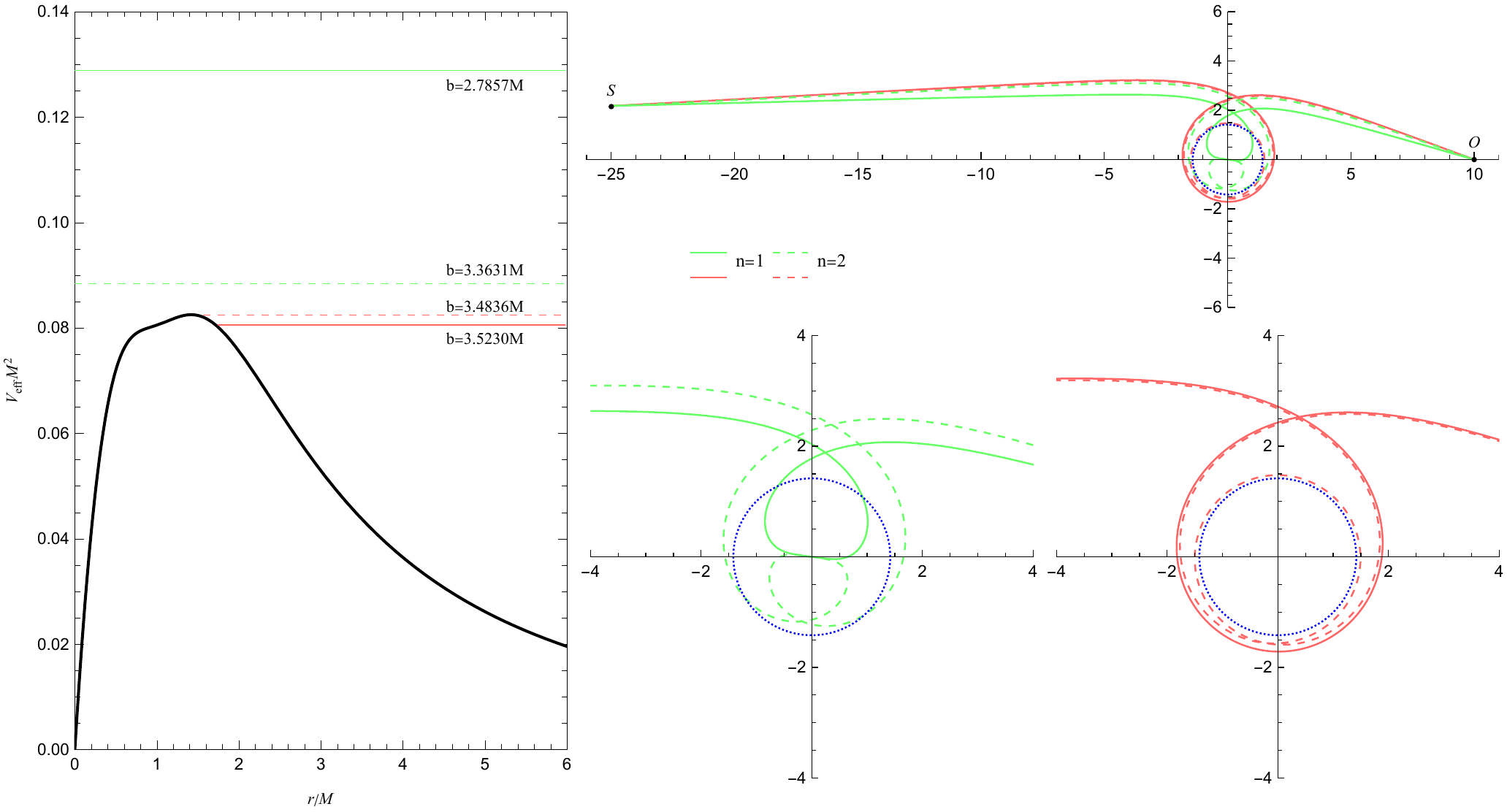}\caption{\textbf{Left}: The
effective potential of a Born-Infeld naked singularity with $\sqrt{P^{2}%
+Q^{2}}/M=1.2$ and $a/M^{2}=2$. It exhibits a single peak corresponding to a
photon sphere at $r_{c}=1.4179M$ with the critical impact parameter
$b_{c}=3.4811M$. The horizontal lines denote $b^{-2}$ of light rays in the
right panel. \textbf{Right}: Light rays connecting the source $S$ with the
observer $O$. The red and green lines represent light rays with $b>b_{c}$ and
$b<b_{c}$, respectively. The light rays with $b<b_{c}$ pass through the
singularity and generate relativistic images inside the critical curve in the
image plane. The blue dashed lines depict the photon sphere, while the solid
and dashed lines illustrate light rays orbiting once and twice around the
photon sphere, respectively.}%
\label{fig: ray-single}%
\end{figure}

We first study strong gravitational lensing in a Born-Infeld naked singularity
with a single photon sphere. When the impact parameter of photons approaches
the critical value, whether from below or above, they undergo significant
deflections. It is important to highlight that photons can traverse the naked
singularity if their impact parameter is smaller than $b_{c}$. Moreover,
photons can orbit the photon sphere in either a clockwise or counterclockwise
direction. Consequently, a distant source yields four relativistic images of
$n^{\text{\text{th}}}$-order, where $n$ is a specified value.

Photons with $b>b_{c}$ reach a turning point $r_{0}$, located just outside the
photon sphere $r_{c}$, as depicted by the red lines in FIG.
\ref{fig: ray-single}. In such cases, the deflection angle in the regime of
strong lensing is described by the equation
\cite{Bozza:2002zj,Tsukamoto:2016jzh}%
\begin{equation}
\alpha\left(  b\right)  =-\bar{a}\ln\left(  b/b_{c}-1\right)  +\bar
{b}+\mathcal{O}\left(  \left(  b/b_{c}-1\right)  \ln\left(  b/b_{c}-1\right)
\right)  ,\label{eq:alpha-a}%
\end{equation}
where
\begin{align}
\bar{a} &  =\sqrt{\frac{2f(r_{c})}{h(r_{c})\left[  R^{\prime\prime}%
(r_{c})f(r_{c})-R(r_{c})f^{\prime\prime}(r_{c})\right]  }},\nonumber\\
\bar{b} &  =\bar{a}\ln\left[  r_{c}^{2}\left(  \frac{R^{\prime\prime}(r_{c}%
)}{R(r_{c})}-\frac{f^{\prime\prime}(r_{c})}{f(r_{c})}\right)  \right]
+I_{R}(r_{c})-\pi.
\end{align}
Here, the term $I_{R}(r_{c})$ represents a regular integral that can be
computed numerically. In the idealized lens model, the angular position of the
image is related to the impact parameter by $b=D_{OL}\vartheta$. By utilizing
the deflection angle formula $\left(  \ref{eq:alpha-a}\right)  $ in
conjunction with the lens equation $\left(  \ref{eq:lens equation}\right)  $,
one can solve for the angular position $\vartheta_{\pm n}^{>}$ for $n^{\text{th}}%
$-order relativistic images produced by photons orbiting the photon sphere $n$
times. It is noteworthy that $-$ and $+$ in the subscript of $\vartheta_{\pm
n}^{>}$ signify counterclockwise and clockwise orbits, respectively, while the
superscript $>$ indicates photons with $b>b_{c}$. Specifically, the angular
position $\vartheta_{\pm n}^{>}$ is given by \cite{Bozza:2002zj}
\begin{equation}
\vartheta_{\pm n}^{>}=\vartheta_{\pm n}^{>0}+\frac{b_{c}e_{n}D_{OS}}{\bar
{a}D_{LS}D_{OL}}\left(  \beta-\vartheta_{\pm n}^{>0}\right)
,\label{eq:theta-out}%
\end{equation}
where $e_{n}=e^{\frac{\bar{b}-2\pi n}{\bar{a}}}$, and $\vartheta_{\pm n}^{>0}%
$, satisfying $\alpha\left(  \vartheta_{\pm n}^{>0}\right)  =\pm2n\pi$,  is
given by%
\begin{equation}
\vartheta_{n}^{>0}=-\vartheta_{-n}^{>0}=\frac{b_{c}}{D_{OL}}\left(
1+e_{n}\right)  .\label{eq:Ering-a}%
\end{equation}

When $b<b_{c}$, photons emitted from the source can pass through the
singularity, resulting in the generation of relativistic images within the
critical curve, as depicted by the green lines in FIG. \ref{fig: ray-single}.
To facilitate the derivation of the integral $\left(  \ref{eq:integral-pass}%
\right)  $, we introduce a variable $z$ defined as
\begin{equation}
z=1-\frac{2r_{c}}{r+r_{c}}.
\end{equation}
The integral $I(b)$ can then be expressed as
\begin{equation}
I(b)=\int_{-1}^{1}A(z)D(z,b)dz,\label{eq:Iexact-b2}%
\end{equation}
where
\begin{equation}
A(z)=\frac{4r_{c}}{(1-z)^{2}R(r)}\sqrt{\frac{R(r_{c})f(r)}{h(r)}},\quad
D(z,b)=\frac{1}{\sqrt{R(r_{c})/b^{2}-f(r)R(r_{c})/R(r)}}.
\end{equation}
Note that $A(z)$ is a regular function of $z$, whereas $D(z,b)$ diverges at
$z=0$ as $b\rightarrow b_{c}$. Hence, we decompose the integral $I(b)$ into a
divergent part $I_{D}(b)$ and a regular part $I_{R}(b)$, as follows,
\begin{align}
I_{D}(b) &  =\int_{-1}^{1}A(0)D_{0}(z,b)dz,\nonumber\\
I_{R}(b) &  =\int_{-1}^{1}\left[  A(z)D\left(  z,b\right)  -A(0)D_{0}%
(z,b_{c})\right]  dz.\label{eq:IR-b2}%
\end{align}
Here, we employ a Taylor expansion within the square root in $D(z,b)$ to
obtain
\begin{equation}
D_{0}(z,b_{c})=\frac{1}{\sqrt{\gamma+\eta z^{2}}},
\end{equation}
where $\gamma$ and $\eta$ are the expansion coefficients. Consequently, the
divergent part $I_{D}(b)$ is given by
\begin{equation}
I_{D}(b)=\frac{A(0)}{\sqrt{\eta}}\ln\left[  \frac{\eta+\sqrt{\eta\left(
\gamma+\eta\right)  }}{-\eta+\sqrt{\eta\left(  \gamma+\eta\right)  }}\right]
.\label{eq:Id}%
\end{equation}
Since the coefficient $\gamma$ approaches zero as $b\rightarrow b_{c}$, the
deflection angle in the strong limit is obtained by expanding $I_{D}(b)$
around $b=b_{c}$,
\begin{equation}
\alpha(b)=-\bar{a}\ln\left(  b_{c}^{2}/b^{2}-1\right)  +\bar{b}+\mathcal{O}%
\left(  \left(  b_{c}/b-1\right)  \ln\left(  b_{c}/b-1\right)  \right)
,\label{eq:alpha-in}%
\end{equation}
where
\begin{align}
\bar{a} &  =2\sqrt{\frac{2f(r_{c})}{h(r_{c})\left[  R^{\prime\prime}%
(r_{c})f(r_{c})-R(r_{c})f^{\prime\prime}(r_{c})\right]  }},\nonumber\\
\bar{b} &  =\bar{a}\ln\left[  8r_{c}^{2}\left(  \frac{R^{\prime\prime}(r_{c}%
)}{R(r_{c})}-\frac{f^{\prime\prime}(r_{c})}{f(r_{c})}\right)  \right]
+I_{R}(b_{c}).
\end{align}
Similarly, the angular position of $n^{\text{th}}$-order relativistic images is given
by
\begin{equation}
\vartheta_{\pm n}^{<}=\vartheta_{\pm n}^{<0}-\frac{b_{c}e_{n}D_{OS}}{2\bar
{a}D_{LS}D_{OL}}\frac{(\beta-\vartheta_{\pm n}^{<0})}{(1+e_{n})^{3/2}%
},\label{eq:theta-in}%
\end{equation}
where the angles $\vartheta_{n}^{<0}$ and $\vartheta_{-n}^{<0}$ are defined
as
\begin{equation}
\vartheta_{n}^{<0}=-\vartheta_{-n}^{<0}=\frac{b_{c}}{D_{OL}}\frac{1}%
{\sqrt{1+e_{n}}}.\label{eq:theta0-in}%
\end{equation}

\begin{table}[ptb]
\begin{tabular}{cccc}
	\hline
	$a/M^{2}$ & 0.2 & 1.5 & 2\tabularnewline
	\hline
	\hline
	$\vartheta_{\pm\infty}$ & $\pm4.1781$ & $\pm15.8189$ & $\pm18.9161$\tabularnewline
	\hline
	$\Delta\vartheta_{1}^{>}$ & $1.0663\times10^{-24}$ & $1.1386$ & $0.2334$\tabularnewline
	$\Delta\vartheta_{2}^{>}$ & $2.1968\times10^{-47}$ & $3.4876\times10^{-7}$ & $1.0560\times10^{-2}$\tabularnewline
	$\Delta\vartheta_{3}^{>}$ & $4.5259\times10^{-70}$ & $1.0682\times10^{-13}$ & $4.7773\times10^{-4}$\tabularnewline
	\hline
	$\Delta\vartheta_{1}^{<}$ & $-3.0217\times10^{-7}$ & $-9.4142$ & $-2.9161$\tabularnewline
	$\Delta\vartheta_{2}^{<}$ & $-1.3715\times10^{-18}$ & $-2.2188\times10^{-2}$ & $-0.7526$\tabularnewline
	$\Delta\vartheta_{3}^{<}$ & $-6.2252\times10^{-30}$ & $-1.2305\times10^{-5}$ & $-0.1679$\tabularnewline
	\hline
	$\Delta\vartheta_{-1}^{>}$ & $-2.8311\times10^{-25}$ & $-0.8133$ & $-0.2179$\tabularnewline
	$\Delta\vartheta_{-2}^{>}$ & $-5.8325\times10^{-48}$ & $-2.4912\times10^{-7}$ & $-9.8579\times10^{-3}$\tabularnewline
	$\Delta\vartheta_{-3}^{>}$ & $-1.2016\times10^{-70}$ & $-7.6304\times10^{-14}$ & $-4.4596\times10^{-4}$\tabularnewline
	\hline
	$\Delta\vartheta_{-1}^{<}$ & $1.6624\times10^{-7}$ & $8.9587$ & $2.8385$\tabularnewline
	$\Delta\vartheta_{-2}^{<}$ & $7.5453\times10^{-19}$ & $1.8780\times10^{-2}$ & $0.7286$\tabularnewline
	$\Delta\vartheta_{-3}^{<}$ & $3.4248\times10^{-30}$ & $1.0412\times10^{-5}$ & $0.1623$\tabularnewline
	\hline
\end{tabular}

\caption{The angular
separation $\Delta\vartheta_{\pm n}^{\gtrless}=\vartheta_{\pm n}^{\gtrless
}-\vartheta_{\pm\infty}^{0}$ between $n^{\text{th}}$-order relativistic images and
the relativistic image formed at the photon sphere in Born-Infeld naked
singularities with a single photon sphere. The parameters $\sqrt{P^{2}+Q^{2}%
}/M=1.2$ and $a/M^{2}=0.2$, $1.5$ and $2$ are considered. The values
$M=4.31\times10^{6}M_{\astrosun}$, $D_{OL}=D_{LS}=7.86$ kpc and $\beta
=2^{\circ}$ are used. The superscripts $>$ and $<$ represent images produced
by light rays with $b>b_{c}$ and $b<b_{c}$, respectively. The subscripts $+n$
and $-n$ indicate images produced by light rays orbiting around the photon
sphere in the clockwise and counterclockwise direction, respectively. All
angles are expressed in units of microarcseconds.}%
\label{table: thetan-single}%
\end{table}

To obtain numerical estimations of $\vartheta_{\pm n}^{\gtrless}$ in an
astrophysical setting, we consider a Born-Infeld naked singularity with
parameters corresponding to the supermassive black hole Sgr A{*} located at
the center of our Galaxy. Specifically, we assume a mass of $M=4.31\times
10^{6}M_{\astrosun}$ and a lens-source distance of $D_{OL}=7.86$ kpc.
Additionally, a source is positioned at $D_{LS}=7.86$ kpc with an angular
separation of $\beta=2^{\circ}$. For Born-Infeld naked singularities with
$\sqrt{P^{2}+Q^{2}}/M=1.2$ and various values of $a/M^{2}$, Table
\ref{table: thetan-single} presents $\Delta\vartheta_{\pm n}^{\gtrless}%
\equiv\vartheta_{\pm n}^{\gtrless}-\vartheta_{\pm\infty}$, where
$\vartheta_{\pm\infty}=\lim\limits_{n\rightarrow\infty}\vartheta_{\pm
n}^{\gtrless0}=\pm b_{c}/D_{OL}$ is the angular position of the relativistic
image formed at the photon sphere. Note that the corresponding effective
potentials of the singularities are displayed in FIGs. \ref{fig:BI} and
\ref{fig: ray-single}. The results demonstrate that as the nonlinear parameter
$a$ increases, the potential peak becomes less pronounced, leading to larger
values of $\Delta\vartheta_{\pm n}^{\gtrless}$. This, in turn, facilitates the
resolution of higher-order relativistic images. Moreover, the relativistic
images with $b<b_{c}$ are more widely separated compared to those with
$b>b_{c}$ due to the significant bending of light rays upon entering or
exiting the photon sphere. Considering a resolution of $0.01$ microarcseconds,
which is capable of resolving the first-order relativistic image in a
Schwarzschild black hole \cite{Bozza:2002zj,Wei:2014dka,Shaikh:2019itn}, it is
observed that relativistic images associated with the singularity having
$a/M^{2}=0.2$ are too closely spaced to be resolved. However, all $n=1$
relativistic images of the singularity with $a/M^{2}=1.5$, as well as $n\leq3$
images of the singularity with $a/M^{2}=2$, can be distinguished.

\subsection{Double Photon Spheres}

\label{subsec:Double-peak potential}

\begin{figure}[ptb]
\includegraphics[width=0.88\textwidth]{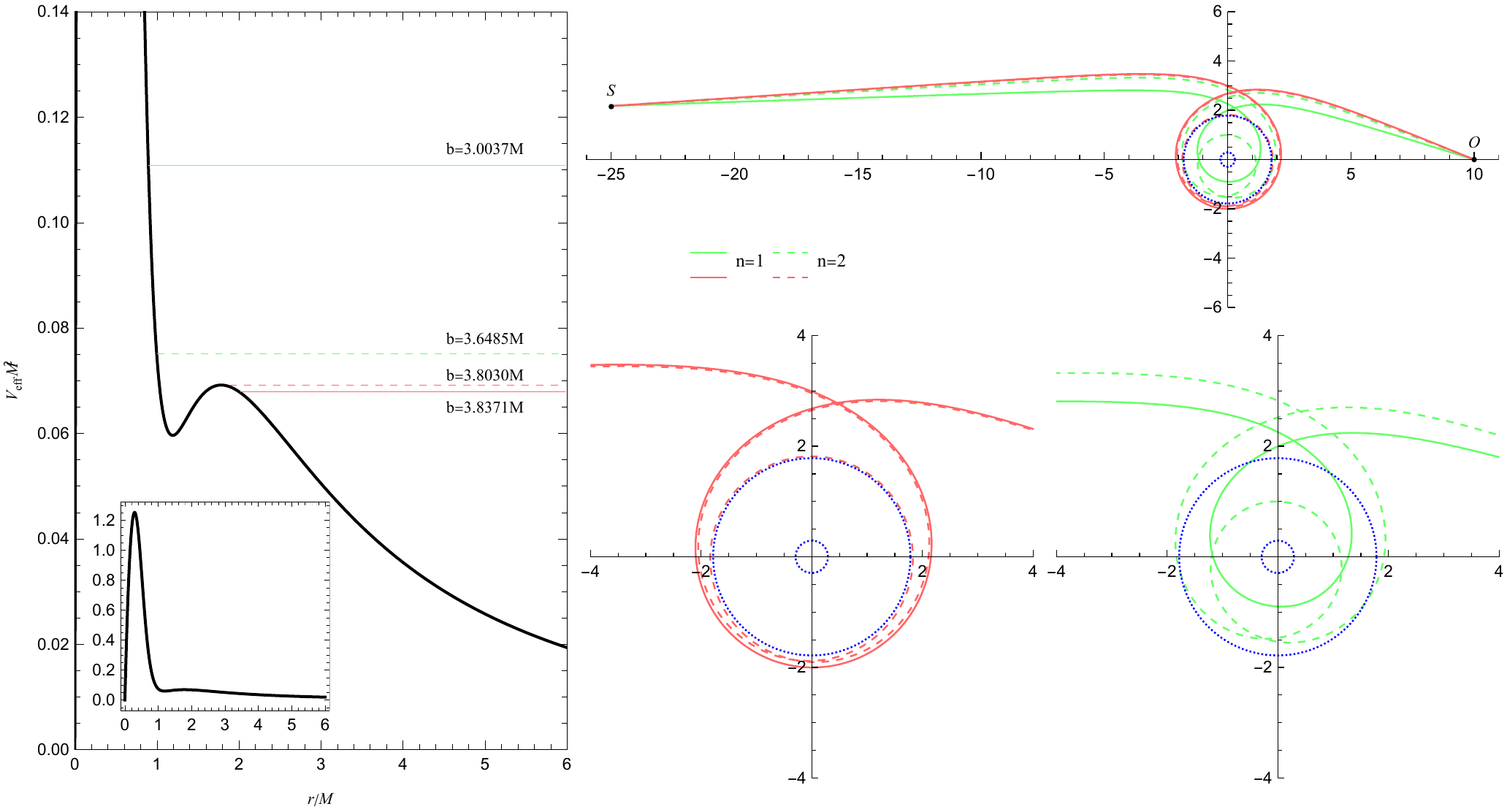}\caption{\textbf{Left}: The
effective potential of a Born-Infeld naked singularity with $\sqrt{P^{2}%
+Q^{2}}/M=1.2$ and $a/M^{2}=0.2$. Notably, it displays two peaks corresponding
to the inner photon sphere at $r_{\text{in}}=0.2919M$ and the outer one at
$r_{\text{out}}=1.7833M$. The horizontal lines denote $b^{-2}$ of light rays
in the right panel, which undergo significant lensing effects near the outer
photon sphere with the impact parameter $b_{\text{out}}=3.8021M$.
\textbf{Right}: Light rays are depicted by red and green lines, indicating
those with $b>b_{\text{out}}$ and $b<b_{\text{out}}$, respectively. The blue
dashed lines represent the photon spheres, while the solid and dashed lines
demonstrate light rays orbiting once and twice around the outer photon sphere,
respectively.}%
\label{fig: ray-double}%
\end{figure}

In the presence of a Born-Infeld naked singularity with a double-peak
effective potential, two photon spheres are observed at distinct locations,
namely, $r=r_{\text{in}}$ and $r=r_{\text{out}}$, with $r_{\text{in}}<$
$r_{\text{out}}$. The critical impact parameters $b_{\text{in}}$ and
$b_{\text{out}}$ represent the impact parameter of light rays on the inner and
outer, respectively. If the height of the inner potential peak is lower than
that of the outer peak, the inner photon sphere remains invisible to a distant
observer, resulting in gravitational lensing similar to the single-peak
scenario. However, when the height of the inner peak surpasses that of the
outer peak, a distant source can generate a total of eight $n^{\text{th}}$-order
relativistic images due to strong gravitational lensing near the inner and
outer photon spheres. The light rays responsible for these relativistic images
are categorized based on their impact parameter $b$,

\begin{itemize}
\item For $b<b_{\text{in}}$, depicted by the green lines in FIG.
\ref{fig: ray-single}, where the potential peak is treated as the inner one.
The light rays emitted from the source pass through the singularity and
produce two relativistic images at $\vartheta=\vartheta_{\pm n}^{\text{in}<}$,
with the minus ($-$) and plus ($+$) signs representing the counterclockwise
and clockwise directions, respectively.

\item For $b>b_{\text{in}}$, illustrated by the red lines in FIG.
\ref{fig: ray-single}. The light rays reach a turning point $r_{0}$ just
outside the inner photon sphere before escaping towards the observer,
generating two relativistic images at $\vartheta=\vartheta_{\pm n}%
^{\text{in}>}$.

\item For $b<b_{\text{out}}$, shown by the green lines in FIG.
\ref{fig: ray-double}. The light rays are reflected at $r=r_{0}$ by the
potential barrier between the two photon spheres, producing two relativistic
images at $\vartheta=\vartheta_{\pm n}^{\text{out}<}$.

\item For $b>b_{\text{out}}$, demonstrated by the red lines in FIG.
\ref{fig: ray-double}. The light rays reach a turning point $r_{0}$ slightly
outside the outer photon sphere, resulting in two relativistic images at
$\vartheta=\vartheta_{\pm n}^{\text{out}>}$.
\end{itemize}

Note that the angular position of the images, $\vartheta_{\pm n}%
^{\text{out}>,\text{in}>}$ and $\vartheta_{\pm n}^{\text{in}<}$, can be
computed using eqns. $\left(  \ref{eq:theta-out}\right)  $ and $\left(
\ref{eq:theta-in}\right)  $, respectively. Moreover, the deflection angle of
light rays with $b<b_{\text{out}}$ has been previously shown to be
\cite{Shaikh:2019itn}%

\begin{equation}
\alpha(b)=-\bar{a}\log\left(  b_{\text{out}}^{2}/b^{2}-1\right)  +\bar
{b}+\mathcal{O}\left(  \left(  b_{\text{out}}/b-1\right)  \ln\left(
b_{\text{out}}/b-1\right)  \right)  ,\label{eq:alpha-in-1}%
\end{equation}
where%
\begin{align}
\bar{a} &  =2\sqrt{\frac{2f(r_{m})}{h(r_{\text{out}})\left[  R^{\prime\prime
}(r_{m})f(r_{m})-R(r_{m})f^{\prime\prime}(r_{m})\right]  }},\nonumber\\
\bar{b} &  =\bar{a}\log\left[  r_{m}^{2}\left(  \frac{r_{m}}{r_{\text{out}}%
}-1\right)  \left(  \frac{R^{\prime\prime}(r_{m})}{R(r_{m})}-\frac
{f^{\prime\prime}(r_{m})}{f(r_{m})}\right)  \right]  +I_{R}(r_{\text{out}%
})-\pi.\label{eq:ab-in-1}%
\end{align}
Here, $r_{m}$ is the critical turning point when $b$ is very close to
$b_{\text{out}}$. Since the divergent parts of the deflection angles $\left(
\ref{eq:alpha-in}\right)  $ and $\left(  \ref{eq:alpha-in-1}\right)  $ share
the same form, one can utilize eqn. $\left(  \ref{eq:theta-in}\right)  $ to
calculate $\vartheta_{\pm n}^{\text{out}<}$ with the values of $\bar{a}$ and
$\bar{b}$ from eqn. $\left(  \ref{eq:ab-in-1}\right)  $.

\begin{table}[ptb]
\begin{tabular}{ccc}
	\hline
	$a/M^{2}$ & 0.2 & 1\tabularnewline
	\hline
	$\vartheta_{\pm\infty}^{\text{in}}$ & $\pm4.8562$ & $\pm17.6478$\tabularnewline
	\hline
	$\Delta\vartheta_{1}^{\text{in}>}$ & $1.0197\times10^{-25}$ & $81.9543$\tabularnewline
	$\Delta\vartheta_{2}^{\text{in}>}$ & $1.4343\times10^{-51}$ & $1.2180\times10^{-20}$\tabularnewline
	$\Delta\vartheta_{3}^{\text{in}>}$ & $2.0176\times10^{-77}$ & $1.8100\times10^{-42}$\tabularnewline
	\hline
	$\Delta\vartheta_{1}^{\text{in}<}$ & $-6.1033\times10^{-7}$ & $-17.6186$\tabularnewline
	$\Delta\vartheta_{2}^{\text{in}<}$ & $-7.2387\times10^{-20}$ & $-3.7064\times10^{-5}$\tabularnewline
	$\Delta\vartheta_{3}^{\text{in}<}$ & $-8.5853\times10^{-33}$ & $-4.5184\times10^{-16}$\tabularnewline
	\hline
	$\Delta\vartheta_{-1}^{\text{in}>}$ & $-2.0781\times10^{-26}$ & $-23.2197$\tabularnewline
	$\Delta\vartheta_{-2}^{\text{in}>}$ & $-2.9232\times10^{-52}$ & $-3.4508\times10^{-21}$\tabularnewline
	$\Delta\vartheta_{-3}^{\text{in}>}$ & $-4.1120\times10^{-78}$ & $-5.1283\times10^{-43}$\tabularnewline
	\hline
	$\Delta\vartheta_{-1}^{\text{in}<}$ & $3.0698\times10^{-7}$ & $17.6091$\tabularnewline
	$\Delta\vartheta_{-2}^{\text{in}<}$ & $3.6408\times10^{-20}$ & $2.0884\times10^{-5}$\tabularnewline
	$\Delta\vartheta_{-3}^{\text{in}<}$ & $4.3181\times10^{-33}$ & $2.5459\times10^{-16}$\tabularnewline
	\hline
\end{tabular}\hspace*{0.5cm}%
\begin{tabular}{cccc}
	\hline
	$a/M^{2}$ & 0.2 & 1 & 1.5\tabularnewline
	\hline
	$\vartheta_{\pm\infty}^{\text{out}}$ & $\pm20.6602$ & $\pm21.4264$ & $\pm21.7967$\tabularnewline
	\hline
	$\Delta\vartheta_{1}^{\text{out}>}$ & $0.2415$ & $0.1491$ & $0.1198$\tabularnewline
	$\Delta\vartheta_{2}^{\text{out}>}$ & $6.5113\times10^{-3}$ & $1.6409\times10^{-3}$ & $9.6643\times10^{-4}$\tabularnewline
	$\Delta\vartheta_{3}^{\text{out}>}$ & $1.7553\times10^{-4}$ & $1.8056\times10^{-5}$ & $7.7952\times10^{-6}$\tabularnewline
	\hline
	$\Delta\vartheta_{1}^{\text{out}<}$ & $-4.6237$ & $-4.6184$ & $-7.5266$\tabularnewline
	$\Delta\vartheta_{2}^{\text{out}<}$ & $-1.0356$ & $-0.6696$ & $-1.1980$\tabularnewline
	$\Delta\vartheta_{3}^{\text{out}<}$ & $-0.1813$ & $-7.3293\times10^{-2}$ & $-0.1162$\tabularnewline
	\hline
	$\Delta\vartheta_{-1}^{\text{out}>}$ & $-0.2229$ & $-0.1349$ & $-0.1076$\tabularnewline
	$\Delta\vartheta_{-2}^{\text{out}>}$ & $-6.0086\times10^{-3}$ & $-1.4843\times10^{-3}$ & $-8.6817\times10^{-4}$\tabularnewline
	$\Delta\vartheta_{-3}^{\text{out}>}$ & $-1.6198\times10^{-4}$ & $-1.6333\times10^{-5}$ & $-7.0027\times10^{-6}$\tabularnewline
	\hline
	$\Delta\vartheta_{-1}^{\text{out}<}$ & $4.4968$ & $4.4581$ & $7.3092$\tabularnewline
	$\Delta\vartheta_{-2}^{\text{out}<}$ & $0.9977$ & $0.6383$ & $1.1403$\tabularnewline
	$\Delta\vartheta_{-3}^{\text{out}<}$ & $0.1743$ & $6.9728\times10^{-2}$ & $0.1102$\tabularnewline
	\hline
\end{tabular}

\caption{Angular separation of
relativistic images near the inner and outer critical curves in Born-Infeld
naked singularities with double photon spheres. The parameters $M$, $D_{OL}$,
$D_{LS}$, $\beta$ and $\sqrt{P^{2}+Q^{2}}/M$ are chosen to be consistent with
those presented in Table \ref{table: thetan-single}. \textbf{Left}: The
angular separation $\Delta\vartheta_{\pm n}^{\text{in}\gtrless}=\vartheta_{\pm
n}^{\text{in}\gtrless}-\vartheta_{\pm\infty}^{\text{in}}$ between $n^{\text{th}}%
$-order relativistic images near the inner critical curve and the relativistic
image formed at the inner photon sphere. \textbf{Right}: The angular
separation $\Delta\vartheta_{\pm n}^{\text{out}\gtrless}=\vartheta_{\pm
n}^{\text{out}\gtrless}-\vartheta_{\pm\infty}^{\text{out}}$ between $n^{\text{th}}%
$-order relativistic images near the outer critical curve and the relativistic
image formed at the outer photon sphere.}%
\label{table: thetan-double}%
\end{table}

Similarly, in the aforementioned astrophysical scenario of Born-Infeld naked
singularities with a single photon sphere, the angular positions of the
relativistic images can be estimated numerically. Specifically, we present the
values of $\Delta\vartheta_{\pm n}^{\text{in}\gtrless}\equiv\vartheta_{\pm
n}^{\text{in}\gtrless}-\vartheta_{\pm\infty}^{\text{in}}$ and $\Delta
\vartheta_{\pm n}^{\text{out}\gtrless}\equiv\vartheta_{\pm n}^{\text{out}%
\gtrless}-\vartheta_{\pm\infty}^{\text{out}}$ for $a/M^{2}=0.2$, $1$ and $1.5$
in Table \ref{table: thetan-double}. Here, $\vartheta_{\pm\infty}^{\text{in}}$
and $\vartheta_{\pm\infty}^{\text{out}}$ are the angular positions of the
relativistic images formed at the inner and outer photon spheres,
respectively. We also assume a resolution of $0.01$ microarcseconds, which
enables the resolution of the first-order image in a Schwarzschild black hole.
For a singularity with $a/M^{2}=0.2$, the inner potential peak is
significantly sharper and higher than the outer peak.  Consequently, the
relativistic images at $\vartheta=\vartheta_{\pm n}^{\text{in}\gtrless}$ are
closely situated near the inner critical curve at $\vartheta=\vartheta
_{\pm\infty}^{\text{in}}$, making them indistinguishable on the image plane.
However, the images near the outer critical curve at $\vartheta=\vartheta
_{\pm\infty}^{\text{out}}$ are well-separated from the outer critical curve,
allowing for the distinction of the first-order images with $b>b_{\text{out}}$
and the first three orders $(n\leq3)$ images with $b<b_{\text{out}}$. When
$a/M^{2}=1$, the inner potential peak becomes flatter, resulting in greater
separation among the images near the inner critical curve. It should be
emphasized that, due to the inner potential peak not being significantly
higher than the outer peak, the formulas for $\vartheta_{\pm n}^{\text{in}>}$
and $\vartheta_{\pm n}^{\text{out}<}$ in the strong deflection limit may
exhibit substantial errors. For the singularity with $a/M^{2}=1.5$, the inner
peak is lower than the outer peak, leading to the existence of relativistic
images solely near the outer critical curve.

\section{Celestial Sphere Images}

\label{sec:Observational Images}

In this section, we investigate gravitational lensing by a Born-Infeld naked
singularity through the image of a luminous celestial sphere centered at the
singularity and surrounding the observers. To obtain the image of the
celestial sphere, we use the numerical backward ray-tracing method to
calculate light rays from the observer to the celestial sphere. Light rays can
be described either by numerically integrating eqn. $\left(  \ref{eq:geo-eq}%
\right)  $ or eight first-order differential equations
\begin{align}
\frac{dx^{\mu}}{d\lambda} &  =p^{\mu},\nonumber\\
\text{ }\frac{dp^{\mu}}{d\lambda} &  =-\tilde{\Gamma}_{\rho\sigma}^{\mu
}p^{\rho}p^{\sigma},\label{eq:eightGeo}%
\end{align}
where $\lambda$ represents the affine parameter, and $\tilde{\Gamma}%
_{\rho\sigma}^{\mu}$ are the Christoffel symbols compatible with the effective
metric. Numerically solving eqn. $\left(  \ref{eq:eightGeo}\right)  $ for
light rays enables us to avoid the need to account for turning points during
the integration, resulting in improved numerical accuracy. Thus, we use eqn.
$\left(  \ref{eq:eightGeo}\right)  $ to calculate the light rays connecting
the observer with the celestial sphere.

For a static observer located at $\left(  t_{o},r_{o},\theta_{o},\varphi
_{o}\right)  $, we introduce a tetrad basis
\begin{equation}
e_{(t)}=\frac{\partial_{t}}{\sqrt{-g_{tt}(r_{o},\theta_{o})}},\text{ }%
e_{(r)}=\frac{\partial_{r}}{\sqrt{g_{rr}(r_{o},\theta_{o})}},\;e_{(\theta
)}=\frac{\partial_{\theta}}{\sqrt{g_{\theta\theta}(r_{o},\theta_{o})}%
},\;e_{(\varphi)}=\frac{\partial_{\varphi}}{\sqrt{g_{\varphi\varphi}%
(r_{o},\theta_{o})}},\label{eq:tetrad}%
\end{equation}
which span the tangent bundle at the observer. To obtain initial conditions
for eqn. $\left(  \ref{eq:eightGeo}\right)  $, a photon captured by the
observer is considered, whose local 4-momentum $\left(  p^{(t)},p^{(r)}%
,p^{(\theta)},p^{(\varphi)}\right)  $ in the tetrad basis is related to the
4-momentum $p_{o}^{\mu}=\left.  dx^{\mu}/d\lambda\right\vert _{\left(
t_{o},r_{o},\theta_{o},\varphi_{o}\right)  }$ as
\begin{equation}
p^{(t)}=\sqrt{f_{\text{BI}}\left(  r_{o}\right)  }p_{o}^{t},\quad
p^{(r)}=p_{o}^{r}/\sqrt{f_{\text{BI}}\left(  r_{o}\right)  },\quad
p^{(\theta)}=r_{o}p_{o}^{\theta},\quad p^{(\varphi)}=r_{o}|\sin\theta
_{o}|p_{o}^{\varphi}.\label{eq:localP}%
\end{equation}
The observation angles $\Theta$ and $\Phi$, as defined in \cite{Cunha:2016bpi}%
, are given by
\begin{equation}
\sin\Theta=\frac{p^{(\theta)}}{p}\text{, }\tan\Phi=\frac{p^{(\varphi)}%
}{p^{(r)}},
\end{equation}
which $p=\sqrt{p^{(r)2}+p^{(\theta)2}+p^{(\varphi)2}}$. We express $p^{(r)}$,
$p^{(\theta)}$ and $p^{(\varphi)}$ in terms of $p$, $\Theta$ and $\Phi$ as
\begin{equation}
p^{\left(  r\right)  }=p\cos\Theta\cos\Phi,\text{ }p^{\left(  \theta\right)
}=p\sin\Theta,\text{ }p^{\left(  \phi\right)  }=p\cos\Theta\sin\Phi.
\end{equation}
Moreover, the condition $\tilde{g}_{\mu\nu}p_{o}^{\mu}p_{o}^{\nu}=0$ and eqn.
$\left(  \ref{eq:localP}\right)  $ give
\begin{equation}
p^{(t)}=p\sqrt{\frac{f_{\text{BI}}\left(  r_{o}\right)  }{f\left(
r_{o}\right)  }}\sqrt{\frac{f_{\text{BI}}\left(  r_{o}\right)  }{h\left(
r_{o}\right)  }\cos^{2}\Theta\cos^{2}\Phi+\frac{R(r_{o})}{r_{o}^{2}}\left(
\sin^{2}\Theta+\sin^{2}\Phi\cos^{2}\Theta\right)  }.
\end{equation}
Using eqn. $\left(  \ref{eq:localP}\right)  $, we can rewrite $p_{o}^{\mu}$ in
terms of $p$, $\Theta$ and $\Phi$, which, together with the coordinates of the
observer, provide initial conditions for eqn. $\left(  \ref{eq:eightGeo}%
\right)  $. Without loss of generality, we set $p=1$ in what follows. The
Cartesian coordinates $\left(  x,y\right)  $ of the image plane of the
observer is defined by
\begin{equation}
x\equiv-r_{\text{o}}\Phi,\text{ }y\equiv r_{\text{o}}\Theta,
\end{equation}
where the sign convention for $\Phi$ leads to the minus sign in the $x$
definition. Note that the direction pointing to the singularity corresponds to
the zero observation angles $\left(  0,0\right)$.

\begin{figure}[ptb]
\includegraphics[width=0.6\textwidth]{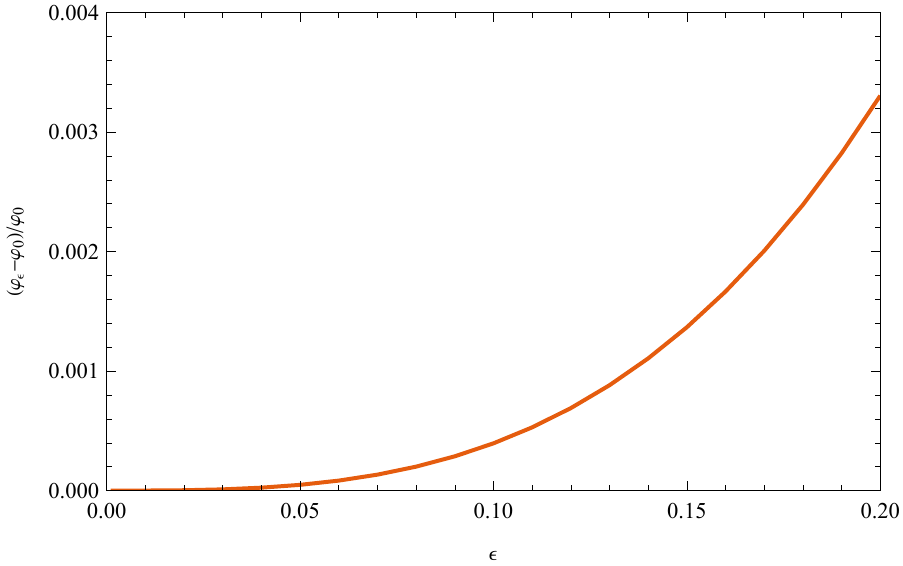}\caption{The relative error
$\left(  \varphi_{\epsilon}-\varphi_{0}\right)  /\varphi_{0}$ as a function of
$\epsilon$ for a light ray emitted from $\left(  r_{e},\varphi_{e}\right)
=\left(  25M,\varphi_{\epsilon}\right)  $ and arriving at $\left(
r_{o},\varphi_{o}\right)  =\left(  10M,\pi\right)  $ with $\left(  \Theta
,\Phi\right)  =\left(  0,3/20\right)  $ on the equatorial plane. The light ray
passes through the singularity, and the ingoing and outgoing branches are
joined at $r=M\epsilon$ during numerical calculations. Here, $\varphi
_{0}=\varphi_{\epsilon=10^{-3}}$, $a/M^{2}=1$ and $\sqrt{Q^{2}+P^{2}}%
/M=1.05$.}%
\label{fig:error}%
\end{figure}

As discussed previously, when a light ray travels through the singularity, its
affine parameter $\lambda$ becomes divergent at $r=0$, posing a challenge for
our numerical implementation. To circumvent this issue, we introduce a small
sphere of radius $M\epsilon$ to enclose the singularity. Outside the sphere,
we can solve eqn. $\left(  \ref{eq:eightGeo}\right)  $ numerically, ensuring
the accuracy and stability of the light ray calculation. Within the sphere, we
can use the expansions in eqn. $\left(  \ref{eq:expansion}\right)  $ to
describe the light ray and provide a connection formula for its entry and exit
points from the sphere. Specifically, we consider the light ray entering and
leaving the sphere at $\left(  t_{\text{in}},r_{\text{in}},\theta_{\text{in}%
},\varphi_{\text{in}}\right)  $ and $\left(  t_{\text{out}},r_{\text{out}%
},\theta_{\text{out}},\varphi_{\text{out}}\right)  $, respectively. With eqns.
$\left(  \ref{eq:asy-solutions}\right)  $ and $\left(  \ref{eq:cc}\right)  $,
one has
\begin{align}
t_{\text{out}} &  =t_{\text{in}}+\mathcal{O}(\epsilon^{2}),\text{
}r_{\text{out}}=r_{\text{in}}=M\epsilon,\text{ }\theta_{\text{out}}=\pi
-\theta_{\text{in}}+\mathcal{O}(\epsilon^{3}),\text{ }\varphi_{\text{out}}%
=\pi+\varphi_{\text{in}}+\mathcal{O}(\epsilon^{3}),\nonumber\\
p_{\text{out}}^{t} &  =p_{\text{in}}^{t},\text{ }p_{\text{out}}^{r}%
=-p_{\text{in}}^{r},\text{ }p_{\text{out}}^{\theta}=-p_{\text{in}}^{\theta
}+\mathcal{O}(\epsilon^{5}\log\epsilon),\text{ }p_{\text{out}}^{\varphi
}=p_{\text{in}}^{\varphi}+\mathcal{O}(\epsilon^{5}\log\epsilon
).\label{eq:inandout}%
\end{align}

In this section, we employ the leading terms of eqn. $\left(
\ref{eq:inandout}\right)  $ to connect the ingoing and outgoing branches. To
explore the numerical error caused by the finite size of $\epsilon$, we
investigate a light ray on the equatorial plane of a Born-Infeld naked
singularity with $a/M^{2}=1$ and $\sqrt{Q^{2}+P^{2}}/M=1.05$. The light ray
originates from $\left(  r_{e},\varphi_{e}\right)  =\left(  25M,\varphi
_{\epsilon}\right)  $, and an observer located at $\left(  r_{o},\varphi
_{o}\right)  =\left(  10M,\pi\right)  $ captures it with observation angles
$\left(  \Theta,\Phi\right)  =\left(  0,3/20\right)  $. To obtain the
coordinate $\varphi_{\epsilon}$, we trace the light ray backward from the
observer to $r_{e}=25M$ while connecting the ingoing branch with the outgoing
one at $r=M\epsilon$. We present the relative error $\left(  \varphi
_{\epsilon}-\varphi_{0}\right)  /\varphi_{0}$ as a function of $\epsilon$ in
FIG. \ref{fig:error}, where $\varphi_{0}\equiv\varphi_{\epsilon=10^{-3}}$. To
maintain numerical precision and efficiency, we set $\epsilon=10^{-1}$ in the
following numerical simulations, for which the relative error is well below
$10^{-3}$.

\begin{figure}[ptb]
\includegraphics[width=0.45\textwidth]{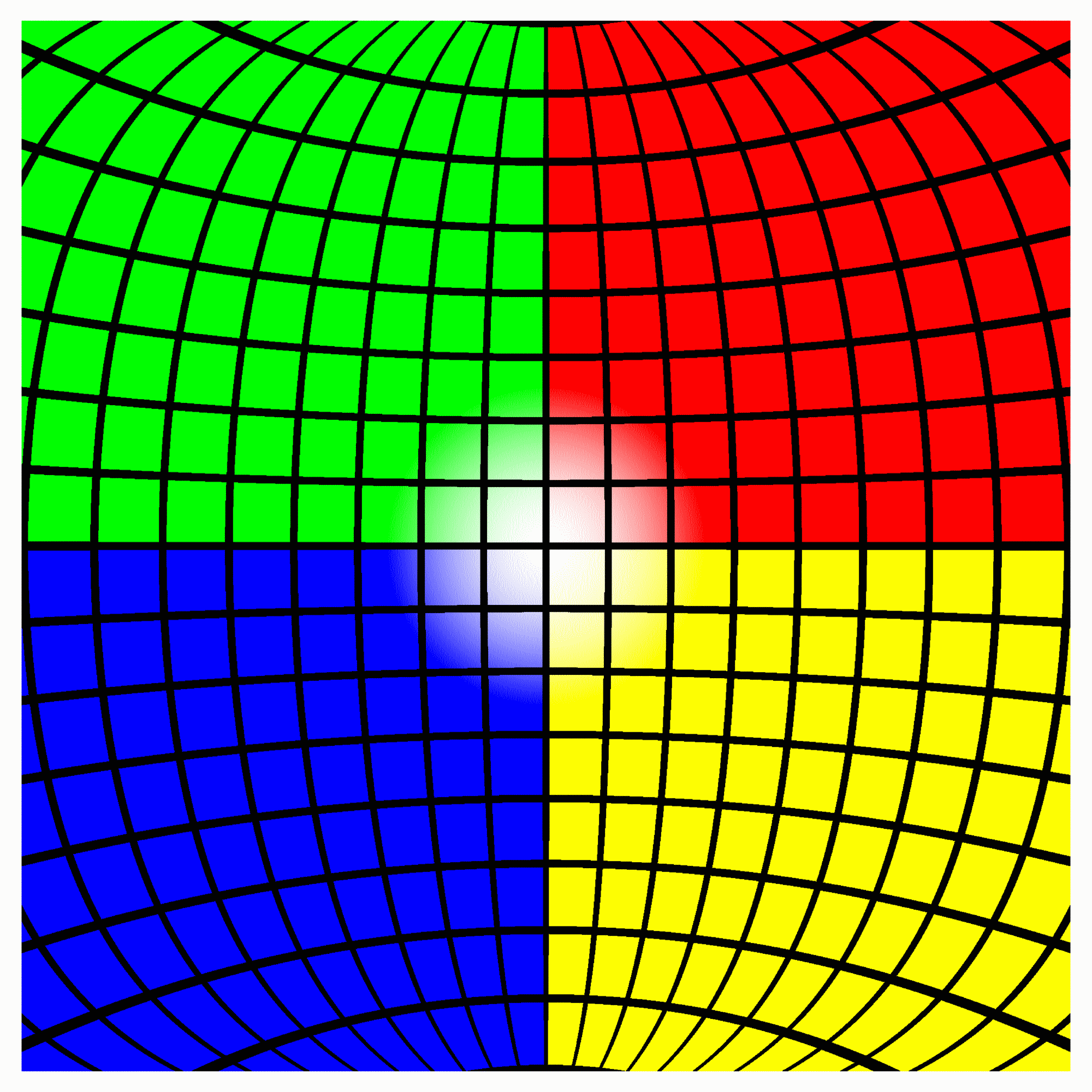}\caption{Observational image
of the celestial sphere in the Minkowski spacetime. The observer is positioned
at $x_{o}^{\mu}=(0,10M,\pi/2,\pi)$ with a field of view of $2\pi/3$.}%
\label{fig: flat and RNNS}%
\end{figure}

To illustrate gravitational lensing by Born-Infeld naked singularities, we
position a luminous celestial sphere at $r_{\text{CS}}=25M$, while an observer
is situated at $x_{o}^{\mu}=(0,10M,\pi/2,\pi)$. The celestial sphere is
divided into four quadrants, each distinguished by a different color, and a
white dot is placed in front of the observer. Additionally, we overlay a grid
of black lines representing constant longitude and latitude, where adjacent
lines are separated by $\pi/18$. To generate an observational image, we vary
the observer's viewing angle and numerically integrate $2000\times2000$ photon
trajectories until they intersect with the celestial sphere. The resulting
image of the celestial sphere in Minkowski spacetime is presented in FIG.
\ref{fig: flat and RNNS}.

\begin{figure}[ptb]
\includegraphics[width=0.45\textwidth]{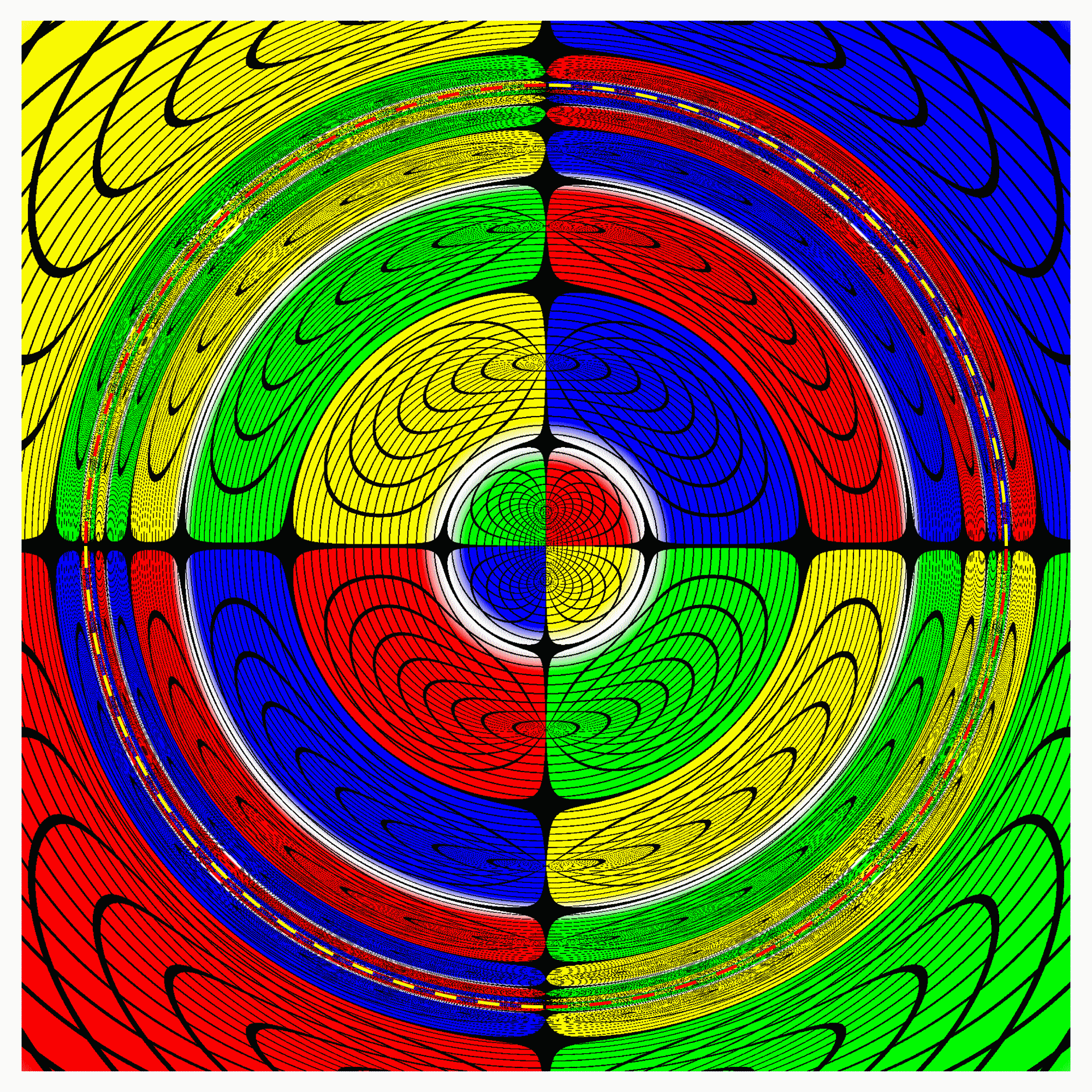}\includegraphics[width=0.45\textwidth]{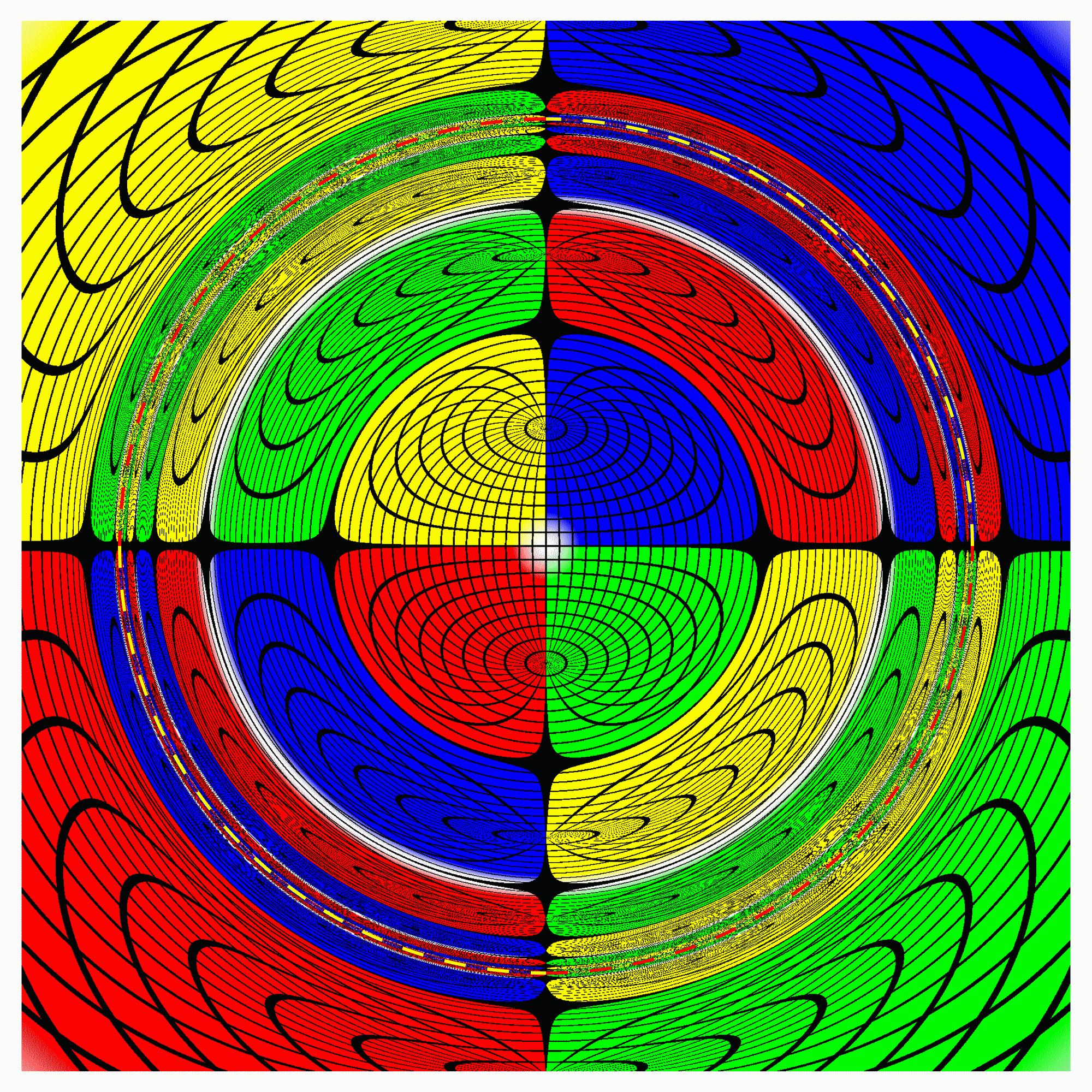}\caption{Images
of the celestial sphere in naked singularities featuring a single photon
sphere. The observer is situated at $x_{o}^{\mu}=(0,10M,\pi/2,\pi)$ with a
field of view of $\pi/4$. The dashed lines depict the critical curve formed by
photons escaping from the photon sphere. \textbf{Left:} The RN naked
singularity with $\sqrt{Q^{2}+P^{2}}/M=1.05$. The image within the critical
curve is generated by light rays that rebound off the infinitely high
potential barrier at the singularity. Three white rings, representing the
Einstein ring of the white dot on the celestial sphere, can be observed within
the critical curve. \textbf{Right:} The Born-Infeld naked singularity with
$\sqrt{Q^{2}+P^{2}}/M=1.2$ and $a/M^{2}=2$. The image within the critical
curve is formed by light rays passing through the singularity. A central white
dot is visible, surrounded by two white rings.}%
\label{fig: Q1.05-I}%
\end{figure}

FIG. \ref{fig: Q1.05-I} displays images of the celestial sphere in both RN and
Born-Infeld naked singularities, both of which possess a single photon sphere.
The dashed circular lines in the images correspond to the critical curve
formed by light rays originating from the photon sphere. Beyond this critical
curve, the images of the celestial sphere in naked singularities bear
resemblance to those observed in black hole spacetime. Notably, unlike shadows
observed in black hole images, the celestial sphere images persist within the
critical curve due to the absence of an event horizon. Additionally,
higher-order images of the celestial sphere can be observed both inside and
outside this critical curve.

The left panel of FIG. \ref{fig: Q1.05-I} illustrates the image of the
celestial sphere in a RN naked singularity characterized by  $\sqrt
{Q^{2}+P^{2}}/M=1.05$. Inside the critical curve, three distinct white rings
can be observed. These rings correspond to the Einstein ring generated by the
white dot positioned on the celestial sphere. Specifically, the innermost
white ring originates from light rays emitted by the white dot, undergoes
reflection at the potential barrier, and eventually reaches the observer after
experiencing an angular coordinate change of $\Delta\varphi=\pi$. Within this
innermost white ring, the reflections from the infinitely high potential
barrier at the singularity produce a mirror image of the celestial sphere.
Moreover, the middle and outermost white rings arise from light rays with
angular coordinate changes of $\Delta\varphi=3\pi$ and $5\pi$, respectively.
In the right panel of FIG. \ref{fig: Q1.05-I}, the image captured in a
Born-Infeld naked singularity with $\sqrt{Q^{2}+P^{2}}/M=1.2$ and $a/M^{2}=2$
is displayed. As expected, light rays passing through the singularity and
undergoing an angular coordinate change of $\Delta\varphi=\pi$ result in a
white dot positioned at the center of the image. Additionally, two white rings
emerge within the critical curve, representing photons that traverse the
singularity with angular coordinate changes of $\Delta\varphi=3\pi$ and $5\pi
$, respectively.

\begin{figure}[ptb]
\includegraphics[width=0.45\textwidth]{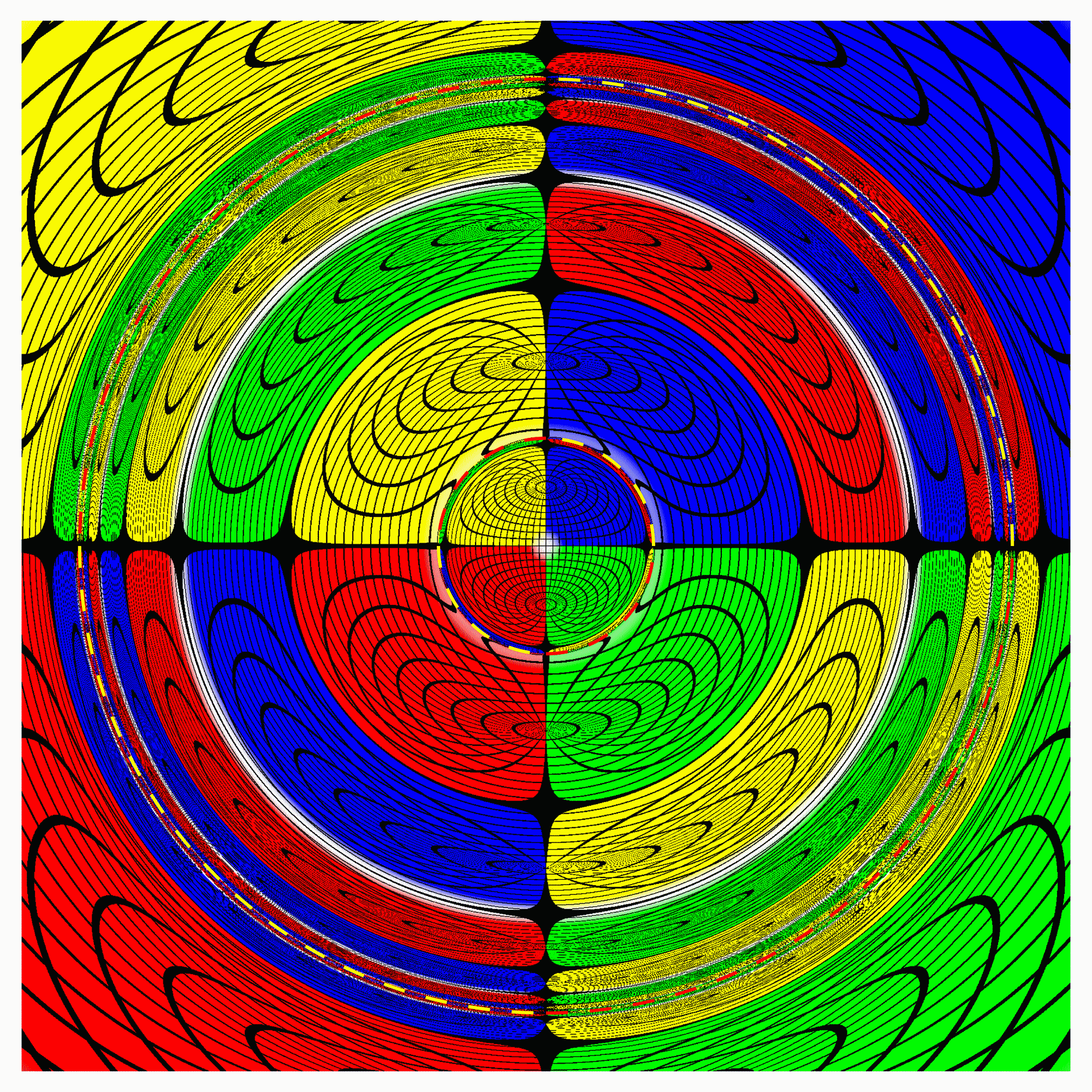}\includegraphics[width=0.45\textwidth]{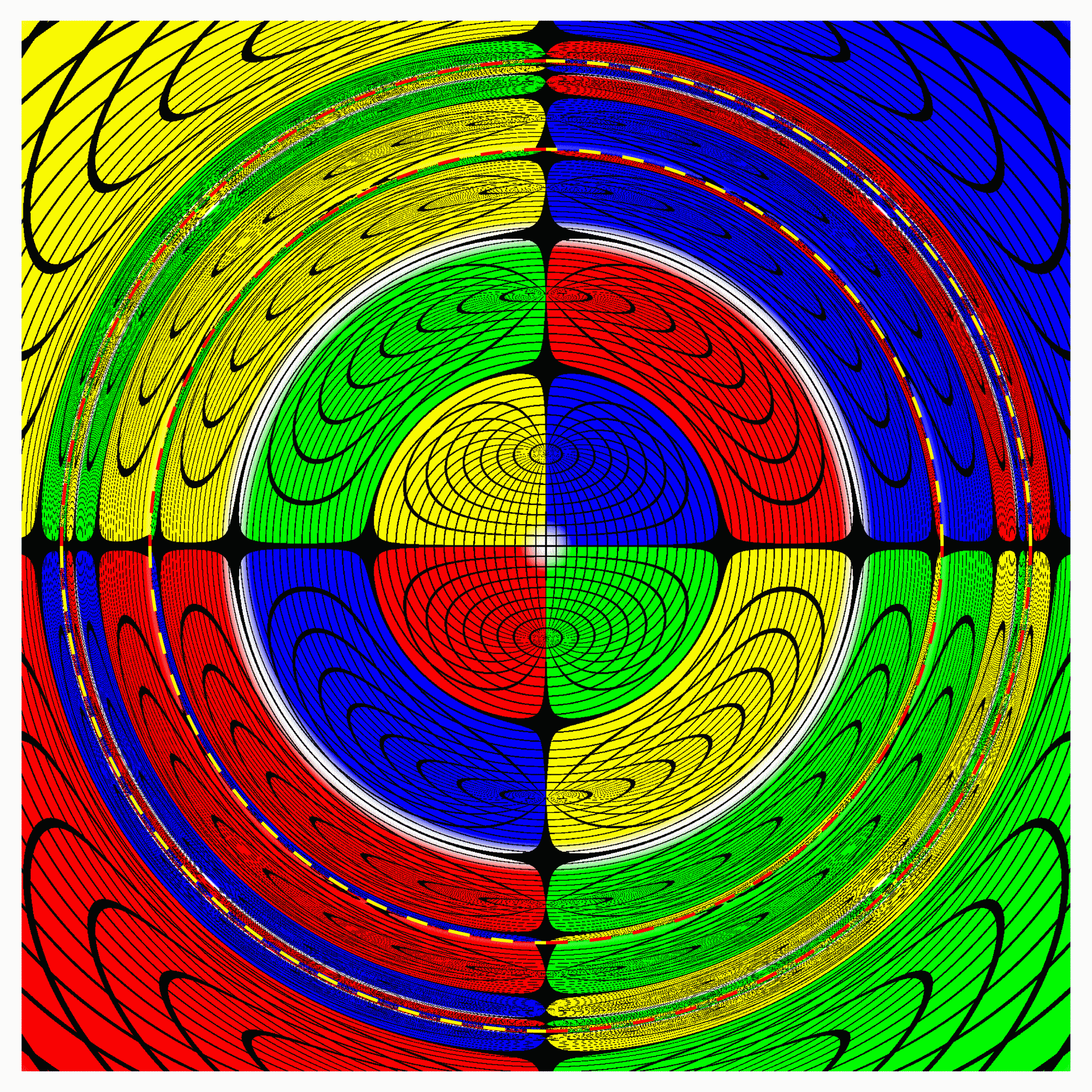}\caption{Images
of the celestial sphere in Born-Infeld naked singularities with $\sqrt
{Q^{2}+P^{2}}/M=1.05$ for $a/M^{2}=0.2$ (\textbf{Left}) and $a/M^{2}=1$
(\textbf{Right}), featuring both inner and outer photon spheres. The observer
is located at $x_{o}^{\mu}=(0,10M,\pi/2,\pi)$ with a field of view of $\pi/4$,
and the corresponding inner and outer critical curves are represented by
dashed lines. The image within the inner critical curve is formed by light
rays traversing the singularity, while the image between the inner and outer
critical curves is a result of light rays reflecting off the potential barrier
situated between the inner and outer potential peaks.}%
\label{fig: Q1.05-II}%
\end{figure}

FIG. \ref{fig: Q1.05-II} showcases images of the celestial sphere in
Born-Infeld naked singularities featuring two photon spheres. The inner and
outer photon spheres give rise to corresponding inner and outer critical
curves, as indicated by the dashed lines in the images. Higher-order celestial
sphere images are observed on both sides of these critical curves. Similarly
to the case of a single photon sphere, a central white spot appears in the
images due to the transparency of the Born-Infeld naked singularity. In the
left panel, characterized by $\sqrt{Q^{2}+P^{2}}/M=1.05$ and $a/M^{2}=0.2$,
two white rings can be observed positioned between the inner and outer
critical curves. These rings originate from photons that are reflected by the
potential barrier located between the inner and outer potential peaks with
$\Delta\varphi=3\pi$ and $5\pi$. In the right panel, with $\sqrt{Q^{2}+P^{2}%
}/M=1.05$ and $a/M^{2}=1.5$, a white ring emerges between the two critical
curves. This ring arises from photons that undergo reflection at the potential
barrier with an angular coordinate change of $\Delta\varphi=5\pi$.
Additionally, a white ring is observed inside the inner critical curve, which
occurs due to photons passing through the singularity and experiencing an
angular coordinate change of $\Delta\varphi=3\pi$.

\section{Conclusions}

\label{sec:CONCLUSIONS}

This paper investigated the phenomenon of gravitational lensing by Born-Infeld
naked singularities, which are solutions of a $\left(  3+1\right)
$-dimensional gravity model coupled to a Born-Infeld electromagnetic field.
Owing to the nonlinearity inherent in Born-Infeld electrodynamics, photons
follow null geodesics of an effective metric, deviating from the background
metric, and remarkably, they are capable of traversing naked singularities.
Additionally, we demonstrated that Born-Infeld naked singularities can exhibit
the presence of two photon spheres with distinct sizes within specific
parameter ranges. The existence of these double photon spheres, combined with
the transparency of naked singularities, significantly impacts the
gravitational lensing of light sources, leading to various effects such as the
emergence of new relativistic images. Consequently, these findings provide a
potent tool for detecting and studying Born-Infeld naked singularities through
their distinctive gravitational lensing signatures.

Naked singularities with double photon spheres have been infrequently
reported; however, it has been discovered that asymptotically flat black holes
can possess two photon spheres outside the event horizon
\cite{Liu:2019rib,Gan:2021pwu,Guo:2022ghl}. A recent investigation focused on
studying the relativistic images produced by point-like light sources and
luminous celestial spheres in the presence of black holes with either a single
or double photon spheres \cite{Guo:2022muy}. The key findings regarding strong
gravitational lensing by black holes and naked singularities can be summarized
as follows:

\begin{itemize}
\item Black holes with a single photon sphere: In the celestial sphere image,
a shadow is observed enclosed by the critical curve, which originates from
light rays escaping the photon sphere. For a point-like source, two $n^{\text{th}}%
$-order relativistic images are present just outside the critical curve,
corresponding to clockwise and counterclockwise winding around the black hole.

\item Black holes with double photon spheres: The celestial sphere image
exhibits both inner and outer critical curves, formed by the inner and outer
photon spheres, respectively. Within the image, there exists a shadow enclosed
by the inner critical curve. Two $n^{\text{th}}$-order relativistic images of a
point-like source appear just outside the outer critical curve, two images are
found just inside the outer critical curve, and two additional images emerge
just outside the inner critical curve.

\item Born-Infeld naked singularities with a single photon sphere: The
celestial sphere image lacks a shadow, and the image within the critical curve
is formed by light rays passing through the singularity. For a point-like
source, there are four $n^{\text{th}}$-order relativistic images, specifically, two
images situated just inside the critical curve and two images positioned just
outside the critical curve.

\item Born-Infeld naked singularities with double photon spheres: The
celestial sphere image does not exhibit a shadow, and the image within the
inner critical curve is produced by light rays that traverse the singularity.
For a point-like source, there are eight $n^{\text{th}}$-order relativistic images,
two images on each side of the inner and outer critical curves.
\end{itemize}

Although current observational facilities lack the capability to distinguish
higher-order relativistic images within the Born-Infeld naked singularity
spacetime, the next-generation Very Long Baseline Interferometry has emerged
as a promising tool for this purpose
\cite{Johnson:2019ljv,Himwich:2020msm,Gralla:2020srx}. Furthermore, it has
been demonstrated that relativistic images located inside the critical curves
are more readily detectable compared to those outside the critical curves.
Hence, it would be highly intriguing if our analysis could be extended to
encompass more astrophysically realistic models, such as the rotating
Born-Infeld naked singularity solution and the imaging of accretion disks.

\begin{acknowledgments}
We are grateful to Qingyu Gan and Xin Jiang for useful discussions and
valuable comments. This work is supported in part by NSFC (Grant No. 12105191, 12275183, 12275184 and 11875196). Houwen Wu is supported by the International Visiting
Program for Excellent Young Scholars of Sichuan University.
\end{acknowledgments}

\bibliographystyle{unsrturl}
\bibliography{ref}

\begin{thebibliography}{100}

\bibitem{Dyson:1920cwa}
F.~W. Dyson, A.~S. Eddington, and C.~Davidson.
\newblock {A Determination of the Deflection of Light by the Sun's
  Gravitational Field, from Observations Made at the Total Eclipse of May 29,
  1919}.
\newblock {\em Phil. Trans. Roy. Soc. Lond. A}, 220:291--333, 1920.
\newblock \href {https://doi.org/10.1098/rsta.1920.0009}
  {\path{doi:10.1098/rsta.1920.0009}}.

\bibitem{Einstein:1936llh}
Albert Einstein.
\newblock {Lens-Like Action of a Star by the Deviation of Light in the
  Gravitational Field}.
\newblock {\em Science}, 84:506--507, 1936.
\newblock \href {https://doi.org/10.1126/science.84.2188.506}
  {\path{doi:10.1126/science.84.2188.506}}.

\bibitem{Eddington:1987tk}
A.~Eddington.
\newblock {\em {SPACE, TIME AND GRAVITATION. AN OUTLINE OF THE GENERAL
  RELATIVITY THEORY}}.
\newblock 1987.

\bibitem{Mellier:1998pk}
Yannick Mellier.
\newblock {Probing the universe with weak lensing}.
\newblock {\em Ann. Rev. Astron. Astrophys.}, 37:127--189, 1999.
\newblock \href {http://arxiv.org/abs/astro-ph/9812172}
  {\path{arXiv:astro-ph/9812172}}, \href
  {https://doi.org/10.1146/annurev.astro.37.1.127}
  {\path{doi:10.1146/annurev.astro.37.1.127}}.

\bibitem{Bartelmann:1999yn}
Matthias Bartelmann and Peter Schneider.
\newblock {Weak gravitational lensing}.
\newblock {\em Phys. Rept.}, 340:291--472, 2001.
\newblock \href {http://arxiv.org/abs/astro-ph/9912508}
  {\path{arXiv:astro-ph/9912508}}, \href
  {https://doi.org/10.1016/S0370-1573(00)00082-X}
  {\path{doi:10.1016/S0370-1573(00)00082-X}}.

\bibitem{Heymans:2013fya}
Catherine Heymans et~al.
\newblock {CFHTLenS tomographic weak lensing cosmological parameter
  constraints: Mitigating the impact of intrinsic galaxy alignments}.
\newblock {\em Mon. Not. Roy. Astron. Soc.}, 432:2433, 2013.
\newblock \href {http://arxiv.org/abs/1303.1808} {\path{arXiv:1303.1808}},
  \href {https://doi.org/10.1093/mnras/stt601}
  {\path{doi:10.1093/mnras/stt601}}.

\bibitem{Kaiser:1992ps}
Nick Kaiser and Gordon Squires.
\newblock {Mapping the dark matter with weak gravitational lensing}.
\newblock {\em Astrophys. J.}, 404:441--450, 1993.
\newblock \href {https://doi.org/10.1086/172297} {\path{doi:10.1086/172297}}.

\bibitem{Clowe:2006eq}
Douglas Clowe, Marusa Bradac, Anthony~H. Gonzalez, Maxim Markevitch, Scott~W.
  Randall, Christine Jones, and Dennis Zaritsky.
\newblock {A direct empirical proof of the existence of dark matter}.
\newblock {\em Astrophys. J. Lett.}, 648:L109--L113, 2006.
\newblock \href {http://arxiv.org/abs/astro-ph/0608407}
  {\path{arXiv:astro-ph/0608407}}, \href {https://doi.org/10.1086/508162}
  {\path{doi:10.1086/508162}}.

\bibitem{Atamurotov:2021hoq}
Farruh Atamurotov, Ahmadjon Abdujabbarov, and Wen-Biao Han.
\newblock {Effect of plasma on gravitational lensing by a Schwarzschild black
  hole immersed in perfect fluid dark matter}.
\newblock {\em Phys. Rev. D}, 104(8):084015, 2021.
\newblock \href {https://doi.org/10.1103/PhysRevD.104.084015}
  {\path{doi:10.1103/PhysRevD.104.084015}}.

\bibitem{Biesiada:2006zf}
Marek Biesiada.
\newblock {Strong lensing systems as a probe of dark energy in the universe}.
\newblock {\em Phys. Rev. D}, 73:023006, 2006.
\newblock \href {https://doi.org/10.1103/PhysRevD.73.023006}
  {\path{doi:10.1103/PhysRevD.73.023006}}.

\bibitem{Cao:2015qja}
Shuo Cao, Marek Biesiada, Rapha Gavazzi, Aleksandra Pi\'orkowska, and Zong-Hong
  Zhu.
\newblock {Cosmology With Strong-lensing Systems}.
\newblock {\em Astrophys. J.}, 806:185, 2015.
\newblock \href {http://arxiv.org/abs/1509.07649} {\path{arXiv:1509.07649}},
  \href {https://doi.org/10.1088/0004-637X/806/2/185}
  {\path{doi:10.1088/0004-637X/806/2/185}}.

\bibitem{DES:2020ahh}
T.~M.~C. Abbott et~al.
\newblock {Dark Energy Survey Year 1 Results: Cosmological constraints from
  cluster abundances and weak lensing}.
\newblock {\em Phys. Rev. D}, 102(2):023509, 2020.
\newblock \href {http://arxiv.org/abs/2002.11124} {\path{arXiv:2002.11124}},
  \href {https://doi.org/10.1103/PhysRevD.102.023509}
  {\path{doi:10.1103/PhysRevD.102.023509}}.

\bibitem{DES:2021wwk}
T.~M.~C. Abbott et~al.
\newblock {Dark Energy Survey Year 3 results: Cosmological constraints from
  galaxy clustering and weak lensing}.
\newblock {\em Phys. Rev. D}, 105(2):023520, 2022.
\newblock \href {http://arxiv.org/abs/2105.13549} {\path{arXiv:2105.13549}},
  \href {https://doi.org/10.1103/PhysRevD.105.023520}
  {\path{doi:10.1103/PhysRevD.105.023520}}.

\bibitem{SDSS:2000jpb}
Xiaohui Fan et~al.
\newblock {The Discovery of a luminous z = 5.80 quasar from the Sloan Digital
  Sky Survey}.
\newblock {\em Astron. J.}, 120:1167--1174, 2000.
\newblock \href {http://arxiv.org/abs/astro-ph/0005414}
  {\path{arXiv:astro-ph/0005414}}, \href {https://doi.org/10.1086/301534}
  {\path{doi:10.1086/301534}}.

\bibitem{Peng:2006ew}
Chien~Y. Peng, Chris~D. Impey, Hans-Walter Rix, Christopher~S. Kochanek,
  Charles~R. Keeton, Emilio~E. Falco, Joseph Lehar, and Brian~A. McLeod.
\newblock {Probing the coevolution of supermassive black holes and galaxies
  using gravitationally lensed quasar hosts}.
\newblock {\em Astrophys. J.}, 649:616--634, 2006.
\newblock \href {http://arxiv.org/abs/astro-ph/0603248}
  {\path{arXiv:astro-ph/0603248}}, \href {https://doi.org/10.1086/506266}
  {\path{doi:10.1086/506266}}.

\bibitem{Oguri:2010ns}
Masamune Oguri and Philip~J. Marshall.
\newblock {Gravitationally lensed quasars and supernovae in future wide-field
  optical imaging surveys}.
\newblock {\em Mon. Not. Roy. Astron. Soc.}, 405:2579--2593, 2010.
\newblock \href {http://arxiv.org/abs/1001.2037} {\path{arXiv:1001.2037}},
  \href {https://doi.org/10.1111/j.1365-2966.2010.16639.x}
  {\path{doi:10.1111/j.1365-2966.2010.16639.x}}.

\bibitem{Yue:2021nwt}
Minghao Yue, Xiaohui Fan, Jinyi Yang, and Feige Wang.
\newblock {Revisiting the Lensed Fraction of High-redshift Quasars}.
\newblock {\em Astrophys. J.}, 925(2):169, 2022.
\newblock \href {http://arxiv.org/abs/2112.02821} {\path{arXiv:2112.02821}},
  \href {https://doi.org/10.3847/1538-4357/ac409b}
  {\path{doi:10.3847/1538-4357/ac409b}}.

\bibitem{Seljak:2003pn}
Uros Seljak and Christopher~M. Hirata.
\newblock {Gravitational lensing as a contaminant of the gravity wave signal in
  CMB}.
\newblock {\em Phys. Rev. D}, 69:043005, 2004.
\newblock \href {http://arxiv.org/abs/astro-ph/0310163}
  {\path{arXiv:astro-ph/0310163}}, \href
  {https://doi.org/10.1103/PhysRevD.69.043005}
  {\path{doi:10.1103/PhysRevD.69.043005}}.

\bibitem{Diego:2021fyd}
Jose~M. Diego, Tom Broadhurst, and George Smoot.
\newblock {Evidence for lensing of gravitational waves from LIGO-Virgo data}.
\newblock {\em Phys. Rev. D}, 104(10):103529, 2021.
\newblock \href {http://arxiv.org/abs/2106.06545} {\path{arXiv:2106.06545}},
  \href {https://doi.org/10.1103/PhysRevD.104.103529}
  {\path{doi:10.1103/PhysRevD.104.103529}}.

\bibitem{Finke:2021znb}
Andreas Finke, Stefano Foffa, Francesco Iacovelli, Michele Maggiore, and
  Michele Mancarella.
\newblock {Probing modified gravitational wave propagation with strongly lensed
  coalescing binaries}.
\newblock {\em Phys. Rev. D}, 104(8):084057, 2021.
\newblock \href {http://arxiv.org/abs/2107.05046} {\path{arXiv:2107.05046}},
  \href {https://doi.org/10.1103/PhysRevD.104.084057}
  {\path{doi:10.1103/PhysRevD.104.084057}}.

\bibitem{Virbhadra:1999nm}
K.~S. Virbhadra and George F.~R. Ellis.
\newblock {Schwarzschild black hole lensing}.
\newblock {\em Phys. Rev. D}, 62:084003, 2000.
\newblock \href {http://arxiv.org/abs/astro-ph/9904193}
  {\path{arXiv:astro-ph/9904193}}, \href
  {https://doi.org/10.1103/PhysRevD.62.084003}
  {\path{doi:10.1103/PhysRevD.62.084003}}.

\bibitem{Akiyama:2019cqa}
Kazunori Akiyama et~al.
\newblock {First M87 Event Horizon Telescope Results. I. The Shadow of the
  Supermassive Black Hole}.
\newblock {\em Astrophys. J. Lett.}, 875:L1, 2019.
\newblock \href {http://arxiv.org/abs/1906.11238} {\path{arXiv:1906.11238}},
  \href {https://doi.org/10.3847/2041-8213/ab0ec7}
  {\path{doi:10.3847/2041-8213/ab0ec7}}.

\bibitem{Akiyama:2019brx}
Kazunori Akiyama et~al.
\newblock {First M87 Event Horizon Telescope Results. II. Array and
  Instrumentation}.
\newblock {\em Astrophys. J. Lett.}, 875(1):L2, 2019.
\newblock \href {http://arxiv.org/abs/1906.11239} {\path{arXiv:1906.11239}},
  \href {https://doi.org/10.3847/2041-8213/ab0c96}
  {\path{doi:10.3847/2041-8213/ab0c96}}.

\bibitem{Akiyama:2019sww}
Kazunori Akiyama et~al.
\newblock {First M87 Event Horizon Telescope Results. III. Data Processing and
  Calibration}.
\newblock {\em Astrophys. J. Lett.}, 875(1):L3, 2019.
\newblock \href {http://arxiv.org/abs/1906.11240} {\path{arXiv:1906.11240}},
  \href {https://doi.org/10.3847/2041-8213/ab0c57}
  {\path{doi:10.3847/2041-8213/ab0c57}}.

\bibitem{Akiyama:2019bqs}
Kazunori Akiyama et~al.
\newblock {First M87 Event Horizon Telescope Results. IV. Imaging the Central
  Supermassive Black Hole}.
\newblock {\em Astrophys. J. Lett.}, 875(1):L4, 2019.
\newblock \href {http://arxiv.org/abs/1906.11241} {\path{arXiv:1906.11241}},
  \href {https://doi.org/10.3847/2041-8213/ab0e85}
  {\path{doi:10.3847/2041-8213/ab0e85}}.

\bibitem{Akiyama:2019fyp}
Kazunori Akiyama et~al.
\newblock {First M87 Event Horizon Telescope Results. V. Physical Origin of the
  Asymmetric Ring}.
\newblock {\em Astrophys. J. Lett.}, 875(1):L5, 2019.
\newblock \href {http://arxiv.org/abs/1906.11242} {\path{arXiv:1906.11242}},
  \href {https://doi.org/10.3847/2041-8213/ab0f43}
  {\path{doi:10.3847/2041-8213/ab0f43}}.

\bibitem{Akiyama:2019eap}
Kazunori Akiyama et~al.
\newblock {First M87 Event Horizon Telescope Results. VI. The Shadow and Mass
  of the Central Black Hole}.
\newblock {\em Astrophys. J. Lett.}, 875(1):L6, 2019.
\newblock \href {http://arxiv.org/abs/1906.11243} {\path{arXiv:1906.11243}},
  \href {https://doi.org/10.3847/2041-8213/ab1141}
  {\path{doi:10.3847/2041-8213/ab1141}}.

\bibitem{Akiyama:2021qum}
Kazunori Akiyama et~al.
\newblock {First M87 Event Horizon Telescope Results. VII. Polarization of the
  Ring}.
\newblock {\em Astrophys. J. Lett.}, 910(1):L12, 2021.
\newblock \href {http://arxiv.org/abs/2105.01169} {\path{arXiv:2105.01169}},
  \href {https://doi.org/10.3847/2041-8213/abe71d}
  {\path{doi:10.3847/2041-8213/abe71d}}.

\bibitem{Akiyama:2021tfw}
Kazunori Akiyama et~al.
\newblock {First M87 Event Horizon Telescope Results. VIII. Magnetic Field
  Structure near The Event Horizon}.
\newblock {\em Astrophys. J. Lett.}, 910(1):L13, 2021.
\newblock \href {http://arxiv.org/abs/2105.01173} {\path{arXiv:2105.01173}},
  \href {https://doi.org/10.3847/2041-8213/abe4de}
  {\path{doi:10.3847/2041-8213/abe4de}}.

\bibitem{EventHorizonTelescope:2022xnr}
Kazunori Akiyama et~al.
\newblock {First Sagittarius A* Event Horizon Telescope Results. I. The Shadow
  of the Supermassive Black Hole in the Center of the Milky Way}.
\newblock {\em Astrophys. J. Lett.}, 930(2):L12, 2022.
\newblock \href {https://doi.org/10.3847/2041-8213/ac6674}
  {\path{doi:10.3847/2041-8213/ac6674}}.

\bibitem{EventHorizonTelescope:2022vjs}
Kazunori Akiyama et~al.
\newblock {First Sagittarius A* Event Horizon Telescope Results. II. EHT and
  Multiwavelength Observations, Data Processing, and Calibration}.
\newblock {\em Astrophys. J. Lett.}, 930(2):L13, 2022.
\newblock \href {https://doi.org/10.3847/2041-8213/ac6675}
  {\path{doi:10.3847/2041-8213/ac6675}}.

\bibitem{EventHorizonTelescope:2022wok}
Kazunori Akiyama et~al.
\newblock {First Sagittarius A* Event Horizon Telescope Results. III. Imaging
  of the Galactic Center Supermassive Black Hole}.
\newblock {\em Astrophys. J. Lett.}, 930(2):L14, 2022.
\newblock \href {https://doi.org/10.3847/2041-8213/ac6429}
  {\path{doi:10.3847/2041-8213/ac6429}}.

\bibitem{EventHorizonTelescope:2022exc}
Kazunori Akiyama et~al.
\newblock {First Sagittarius A* Event Horizon Telescope Results. IV.
  Variability, Morphology, and Black Hole Mass}.
\newblock {\em Astrophys. J. Lett.}, 930(2):L15, 2022.
\newblock \href {https://doi.org/10.3847/2041-8213/ac6736}
  {\path{doi:10.3847/2041-8213/ac6736}}.

\bibitem{EventHorizonTelescope:2022urf}
Kazunori Akiyama et~al.
\newblock {First Sagittarius A* Event Horizon Telescope Results. V. Testing
  Astrophysical Models of the Galactic Center Black Hole}.
\newblock {\em Astrophys. J. Lett.}, 930(2):L16, 2022.
\newblock \href {https://doi.org/10.3847/2041-8213/ac6672}
  {\path{doi:10.3847/2041-8213/ac6672}}.

\bibitem{EventHorizonTelescope:2022xqj}
Kazunori Akiyama et~al.
\newblock {First Sagittarius A* Event Horizon Telescope Results. VI. Testing
  the Black Hole Metric}.
\newblock {\em Astrophys. J. Lett.}, 930(2):L17, 2022.
\newblock \href {https://doi.org/10.3847/2041-8213/ac6756}
  {\path{doi:10.3847/2041-8213/ac6756}}.

\bibitem{Falcke:1999pj}
Heino Falcke, Fulvio Melia, and Eric Agol.
\newblock {Viewing the shadow of the black hole at the galactic center}.
\newblock {\em Astrophys. J. Lett.}, 528:L13, 2000.
\newblock \href {http://arxiv.org/abs/astro-ph/9912263}
  {\path{arXiv:astro-ph/9912263}}, \href {https://doi.org/10.1086/312423}
  {\path{doi:10.1086/312423}}.

\bibitem{Claudel:2000yi}
Clarissa-Marie Claudel, K.~S. Virbhadra, and G.~F.~R. Ellis.
\newblock {The Geometry of photon surfaces}.
\newblock {\em J. Math. Phys.}, 42:818--838, 2001.
\newblock \href {http://arxiv.org/abs/gr-qc/0005050}
  {\path{arXiv:gr-qc/0005050}}, \href {https://doi.org/10.1063/1.1308507}
  {\path{doi:10.1063/1.1308507}}.

\bibitem{Eiroa:2002mk}
Ernesto~F. Eiroa, Gustavo~E. Romero, and Diego~F. Torres.
\newblock {Reissner-Nordstrom black hole lensing}.
\newblock {\em Phys. Rev. D}, 66:024010, 2002.
\newblock \href {http://arxiv.org/abs/gr-qc/0203049}
  {\path{arXiv:gr-qc/0203049}}, \href
  {https://doi.org/10.1103/PhysRevD.66.024010}
  {\path{doi:10.1103/PhysRevD.66.024010}}.

\bibitem{Virbhadra:2008ws}
K.~S. Virbhadra.
\newblock {Relativistic images of Schwarzschild black hole lensing}.
\newblock {\em Phys. Rev. D}, 79:083004, 2009.
\newblock \href {http://arxiv.org/abs/0810.2109} {\path{arXiv:0810.2109}},
  \href {https://doi.org/10.1103/PhysRevD.79.083004}
  {\path{doi:10.1103/PhysRevD.79.083004}}.

\bibitem{Yumoto:2012kz}
Akifumi Yumoto, Daisuke Nitta, Takeshi Chiba, and Naoshi Sugiyama.
\newblock {Shadows of Multi-Black Holes: Analytic Exploration}.
\newblock {\em Phys. Rev. D}, 86:103001, 2012.
\newblock \href {http://arxiv.org/abs/1208.0635} {\path{arXiv:1208.0635}},
  \href {https://doi.org/10.1103/PhysRevD.86.103001}
  {\path{doi:10.1103/PhysRevD.86.103001}}.

\bibitem{Wei:2013kza}
Shao-Wen Wei and Yu-Xiao Liu.
\newblock {Observing the shadow of Einstein-Maxwell-Dilaton-Axion black hole}.
\newblock {\em JCAP}, 11:063, 2013.
\newblock \href {http://arxiv.org/abs/1311.4251} {\path{arXiv:1311.4251}},
  \href {https://doi.org/10.1088/1475-7516/2013/11/063}
  {\path{doi:10.1088/1475-7516/2013/11/063}}.

\bibitem{Zakharov:2014lqa}
Alexander~F. Zakharov.
\newblock {Constraints on a charge in the Reissner-Nordstr\"om metric for the
  black hole at the Galactic Center}.
\newblock {\em Phys. Rev. D}, 90(6):062007, 2014.
\newblock \href {http://arxiv.org/abs/1407.7457} {\path{arXiv:1407.7457}},
  \href {https://doi.org/10.1103/PhysRevD.90.062007}
  {\path{doi:10.1103/PhysRevD.90.062007}}.

\bibitem{Atamurotov:2015xfa}
Farruh Atamurotov, Sushant~G. Ghosh, and Bobomurat Ahmedov.
\newblock {Horizon structure of rotating
  Einstein\textendash{}Born\textendash{}Infeld black holes and shadow}.
\newblock {\em Eur. Phys. J. C}, 76(5):273, 2016.
\newblock \href {http://arxiv.org/abs/1506.03690} {\path{arXiv:1506.03690}},
  \href {https://doi.org/10.1140/epjc/s10052-016-4122-9}
  {\path{doi:10.1140/epjc/s10052-016-4122-9}}.

\bibitem{Cunha:2016wzk}
Pedro V.~P. Cunha, Carlos A.~R. Herdeiro, Burkhard Kleihaus, Jutta Kunz, and
  Eugen Radu.
\newblock {Shadows of
  Einstein\textendash{}dilaton\textendash{}Gauss\textendash{}Bonnet black
  holes}.
\newblock {\em Phys. Lett. B}, 768:373--379, 2017.
\newblock \href {http://arxiv.org/abs/1701.00079} {\path{arXiv:1701.00079}},
  \href {https://doi.org/10.1016/j.physletb.2017.03.020}
  {\path{doi:10.1016/j.physletb.2017.03.020}}.

\bibitem{Dastan:2016bfy}
Sara Dastan, Reza Saffari, and Saheb Soroushfar.
\newblock {Shadow of a Kerr-Sen dilaton-axion Black Hole}.
\newblock 10 2016.
\newblock \href {http://arxiv.org/abs/1610.09477} {\path{arXiv:1610.09477}}.

\bibitem{Amir:2017slq}
Muhammed Amir, Balendra~Pratap Singh, and Sushant~G. Ghosh.
\newblock {Shadows of rotating five-dimensional charged EMCS black holes}.
\newblock {\em Eur. Phys. J. C}, 78(5):399, 2018.
\newblock \href {http://arxiv.org/abs/1707.09521} {\path{arXiv:1707.09521}},
  \href {https://doi.org/10.1140/epjc/s10052-018-5872-3}
  {\path{doi:10.1140/epjc/s10052-018-5872-3}}.

\bibitem{Wang:2017hjl}
Mingzhi Wang, Songbai Chen, and Jiliang Jing.
\newblock {Shadow casted by a Konoplya-Zhidenko rotating non-Kerr black hole}.
\newblock {\em JCAP}, 10:051, 2017.
\newblock \href {http://arxiv.org/abs/1707.09451} {\path{arXiv:1707.09451}},
  \href {https://doi.org/10.1088/1475-7516/2017/10/051}
  {\path{doi:10.1088/1475-7516/2017/10/051}}.

\bibitem{Ovgun:2018tua}
Ali \"Ovg\"un, \.Izzet Sakall\i{}, and Joel Saavedra.
\newblock {Shadow cast and Deflection angle of Kerr-Newman-Kasuya spacetime}.
\newblock {\em JCAP}, 10:041, 2018.
\newblock \href {http://arxiv.org/abs/1807.00388} {\path{arXiv:1807.00388}},
  \href {https://doi.org/10.1088/1475-7516/2018/10/041}
  {\path{doi:10.1088/1475-7516/2018/10/041}}.

\bibitem{Perlick:2018iye}
Volker Perlick, Oleg~Yu. Tsupko, and Gennady~S. Bisnovatyi-Kogan.
\newblock {Black hole shadow in an expanding universe with a cosmological
  constant}.
\newblock {\em Phys. Rev. D}, 97(10):104062, 2018.
\newblock \href {http://arxiv.org/abs/1804.04898} {\path{arXiv:1804.04898}},
  \href {https://doi.org/10.1103/PhysRevD.97.104062}
  {\path{doi:10.1103/PhysRevD.97.104062}}.

\bibitem{Kumar:2019pjp}
Rahul Kumar, Sushant~G. Ghosh, and Anzhong Wang.
\newblock {Shadow cast and deflection of light by charged rotating regular
  black holes}.
\newblock {\em Phys. Rev. D}, 100(12):124024, 2019.
\newblock \href {http://arxiv.org/abs/1912.05154} {\path{arXiv:1912.05154}},
  \href {https://doi.org/10.1103/PhysRevD.100.124024}
  {\path{doi:10.1103/PhysRevD.100.124024}}.

\bibitem{Zhu:2019ura}
Tao Zhu, Qiang Wu, Mubasher Jamil, and Kimet Jusufi.
\newblock {Shadows and deflection angle of charged and slowly rotating black
  holes in Einstein-\AE{}ther theory}.
\newblock {\em Phys. Rev. D}, 100(4):044055, 2019.
\newblock \href {http://arxiv.org/abs/1906.05673} {\path{arXiv:1906.05673}},
  \href {https://doi.org/10.1103/PhysRevD.100.044055}
  {\path{doi:10.1103/PhysRevD.100.044055}}.

\bibitem{Ma:2019ybz}
Liang Ma and H.~Lu.
\newblock {Bounds on photon spheres and shadows of charged black holes in
  Einstein-Gauss-Bonnet-Maxwell gravity}.
\newblock {\em Phys. Lett. B}, 807:135535, 2020.
\newblock \href {http://arxiv.org/abs/1912.05569} {\path{arXiv:1912.05569}},
  \href {https://doi.org/10.1016/j.physletb.2020.135535}
  {\path{doi:10.1016/j.physletb.2020.135535}}.

\bibitem{Mishra:2019trb}
Akash~K. Mishra, Sumanta Chakraborty, and Sudipta Sarkar.
\newblock {Understanding photon sphere and black hole shadow in dynamically
  evolving spacetimes}.
\newblock {\em Phys. Rev. D}, 99(10):104080, 2019.
\newblock \href {http://arxiv.org/abs/1903.06376} {\path{arXiv:1903.06376}},
  \href {https://doi.org/10.1103/PhysRevD.99.104080}
  {\path{doi:10.1103/PhysRevD.99.104080}}.

\bibitem{Zeng:2020dco}
Xiao-Xiong Zeng, Hai-Qing Zhang, and Hongbao Zhang.
\newblock {Shadows and photon spheres with spherical accretions in the
  four-dimensional Gauss\textendash{}Bonnet black hole}.
\newblock {\em Eur. Phys. J. C}, 80(9):872, 2020.
\newblock \href {http://arxiv.org/abs/2004.12074} {\path{arXiv:2004.12074}},
  \href {https://doi.org/10.1140/epjc/s10052-020-08449-y}
  {\path{doi:10.1140/epjc/s10052-020-08449-y}}.

\bibitem{Zeng:2020vsj}
Xiao-Xiong Zeng and Hai-Qing Zhang.
\newblock {Influence of quintessence dark energy on the shadow of black hole}.
\newblock {\em Eur. Phys. J. C}, 80(11):1058, 2020.
\newblock \href {http://arxiv.org/abs/2007.06333} {\path{arXiv:2007.06333}},
  \href {https://doi.org/10.1140/epjc/s10052-020-08656-7}
  {\path{doi:10.1140/epjc/s10052-020-08656-7}}.

\bibitem{Saurabh:2020zqg}
K.~Saurabh and Kimet Jusufi.
\newblock {Imprints of dark matter on black hole shadows using spherical
  accretions}.
\newblock {\em Eur. Phys. J. C}, 81(6):490, 2021.
\newblock \href {http://arxiv.org/abs/2009.10599} {\path{arXiv:2009.10599}},
  \href {https://doi.org/10.1140/epjc/s10052-021-09280-9}
  {\path{doi:10.1140/epjc/s10052-021-09280-9}}.

\bibitem{Roy:2020dyy}
Rittick Roy and Sayan Chakrabarti.
\newblock {Study on black hole shadows in asymptotically de Sitter spacetimes}.
\newblock {\em Phys. Rev. D}, 102(2):024059, 2020.
\newblock \href {http://arxiv.org/abs/2003.14107} {\path{arXiv:2003.14107}},
  \href {https://doi.org/10.1103/PhysRevD.102.024059}
  {\path{doi:10.1103/PhysRevD.102.024059}}.

\bibitem{Li:2020drn}
Peng-Cheng Li, Minyong Guo, and Bin Chen.
\newblock {Shadow of a Spinning Black Hole in an Expanding Universe}.
\newblock {\em Phys. Rev. D}, 101(8):084041, 2020.
\newblock \href {http://arxiv.org/abs/2001.04231} {\path{arXiv:2001.04231}},
  \href {https://doi.org/10.1103/PhysRevD.101.084041}
  {\path{doi:10.1103/PhysRevD.101.084041}}.

\bibitem{Kumar:2020hgm}
Rahul Kumar, Sushant~G. Ghosh, and Anzhong Wang.
\newblock {Gravitational deflection of light and shadow cast by rotating
  Kalb-Ramond black holes}.
\newblock {\em Phys. Rev. D}, 101(10):104001, 2020.
\newblock \href {http://arxiv.org/abs/2001.00460} {\path{arXiv:2001.00460}},
  \href {https://doi.org/10.1103/PhysRevD.101.104001}
  {\path{doi:10.1103/PhysRevD.101.104001}}.

\bibitem{Zhang:2020xub}
Ming Zhang and Jie Jiang.
\newblock {Shadows of accelerating black holes}.
\newblock {\em Phys. Rev. D}, 103(2):025005, 2021.
\newblock \href {http://arxiv.org/abs/2010.12194} {\path{arXiv:2010.12194}},
  \href {https://doi.org/10.1103/PhysRevD.103.025005}
  {\path{doi:10.1103/PhysRevD.103.025005}}.

\bibitem{Guerrero:2022qkh}
Merce Guerrero, Gonzalo~J. Olmo, Diego Rubiera-Garcia, and Diego G\'omez
  S\'aez-Chill\'on.
\newblock {Light ring images of double photon spheres in black hole and
  wormhole spacetimes}.
\newblock {\em Phys. Rev. D}, 105(8):084057, 2022.
\newblock \href {http://arxiv.org/abs/2202.03809} {\path{arXiv:2202.03809}},
  \href {https://doi.org/10.1103/PhysRevD.105.084057}
  {\path{doi:10.1103/PhysRevD.105.084057}}.

\bibitem{Virbhadra:2022iiy}
K.~S. Virbhadra.
\newblock {Distortions of images of Schwarzschild lensing}.
\newblock 4 2022.
\newblock \href {http://arxiv.org/abs/2204.01879} {\path{arXiv:2204.01879}}.

\bibitem{Schmidt:2008hc}
Fabian Schmidt.
\newblock {Weak Lensing Probes of Modified Gravity}.
\newblock {\em Phys. Rev. D}, 78:043002, 2008.
\newblock \href {http://arxiv.org/abs/0805.4812} {\path{arXiv:0805.4812}},
  \href {https://doi.org/10.1103/PhysRevD.78.043002}
  {\path{doi:10.1103/PhysRevD.78.043002}}.

\bibitem{Guzik:2009cm}
Jacek Guzik, Bhuvnesh Jain, and Masahiro Takada.
\newblock {Tests of Gravity from Imaging and Spectroscopic Surveys}.
\newblock {\em Phys. Rev. D}, 81:023503, 2010.
\newblock \href {http://arxiv.org/abs/0906.2221} {\path{arXiv:0906.2221}},
  \href {https://doi.org/10.1103/PhysRevD.81.023503}
  {\path{doi:10.1103/PhysRevD.81.023503}}.

\bibitem{Liao:2015uzb}
Kai Liao, Zhengxiang Li, Shuo Cao, Marek Biesiada, Xiaogang Zheng, and
  Zong-Hong Zhu.
\newblock {The Distance Duality Relation From Strong Gravitational Lensing}.
\newblock {\em Astrophys. J.}, 822(2):74, 2016.
\newblock \href {http://arxiv.org/abs/1511.01318} {\path{arXiv:1511.01318}},
  \href {https://doi.org/10.3847/0004-637X/822/2/74}
  {\path{doi:10.3847/0004-637X/822/2/74}}.

\bibitem{Goulart:2017iko}
Prieslei Goulart.
\newblock {Phantom wormholes in Einstein\textendash{}Maxwell-dilaton theory}.
\newblock {\em Class. Quant. Grav.}, 35(2):025012, 2018.
\newblock \href {http://arxiv.org/abs/1708.00935} {\path{arXiv:1708.00935}},
  \href {https://doi.org/10.1088/1361-6382/aa9dfc}
  {\path{doi:10.1088/1361-6382/aa9dfc}}.

\bibitem{Nascimento:2020ime}
J.~R. Nascimento, A.~Yu. Petrov, P.~J. Porfirio, and A.~R. Soares.
\newblock {Gravitational lensing in black-bounce spacetimes}.
\newblock {\em Phys. Rev. D}, 102(4):044021, 2020.
\newblock \href {http://arxiv.org/abs/2005.13096} {\path{arXiv:2005.13096}},
  \href {https://doi.org/10.1103/PhysRevD.102.044021}
  {\path{doi:10.1103/PhysRevD.102.044021}}.

\bibitem{Qin:2020xzu}
Xin Qin, Songbai Chen, and Jiliang Jing.
\newblock {Image of a regular phantom compact object and its luminosity under
  spherical accretions}.
\newblock {\em Class. Quant. Grav.}, 38(11):115008, 2021.
\newblock \href {http://arxiv.org/abs/2011.04310} {\path{arXiv:2011.04310}},
  \href {https://doi.org/10.1088/1361-6382/abf712}
  {\path{doi:10.1088/1361-6382/abf712}}.

\bibitem{Junior:2021svb}
Haroldo C. D.~Lima Junior, Jian-Zhi Yang, Lu\'\i{}s C.~B. Crispino, Pedro V.~P.
  Cunha, and Carlos A.~R. Herdeiro.
\newblock {Einstein-Maxwell-dilaton neutral black holes in strong magnetic
  fields: Topological charge, shadows, and lensing}.
\newblock {\em Phys. Rev. D}, 105(6):064070, 2022.
\newblock \href {http://arxiv.org/abs/2112.10802} {\path{arXiv:2112.10802}},
  \href {https://doi.org/10.1103/PhysRevD.105.064070}
  {\path{doi:10.1103/PhysRevD.105.064070}}.

\bibitem{Islam:2021ful}
Shafqat~Ul Islam, Jitendra Kumar, and Sushant~G. Ghosh.
\newblock {Strong gravitational lensing by rotating Simpson-Visser black
  holes}.
\newblock {\em JCAP}, 10:013, 2021.
\newblock \href {http://arxiv.org/abs/2104.00696} {\path{arXiv:2104.00696}},
  \href {https://doi.org/10.1088/1475-7516/2021/10/013}
  {\path{doi:10.1088/1475-7516/2021/10/013}}.

\bibitem{Tsukamoto:2021caq}
Naoki Tsukamoto.
\newblock {Gravitational lensing by two photon spheres in a black-bounce
  spacetime in strong deflection limits}.
\newblock 5 2021.
\newblock \href {http://arxiv.org/abs/2105.14336} {\path{arXiv:2105.14336}}.

\bibitem{Olmo:2021piq}
Gonzalo~J. Olmo, Diego Rubiera-Garcia, and Diego S\'aez-Chill\'on G\'omez.
\newblock {New light rings from multiple critical curves as observational
  signatures of black hole mimickers}.
\newblock {\em Phys. Lett. B}, 829:137045, 2022.
\newblock \href {http://arxiv.org/abs/2110.10002} {\path{arXiv:2110.10002}},
  \href {https://doi.org/10.1016/j.physletb.2022.137045}
  {\path{doi:10.1016/j.physletb.2022.137045}}.

\bibitem{Shapiro:1991zza}
Stuart~L. Shapiro and Saul~A. Teukolsky.
\newblock {Formation of naked singularities: The violation of cosmic
  censorship}.
\newblock {\em Phys. Rev. Lett.}, 66:994--997, 1991.
\newblock \href {https://doi.org/10.1103/PhysRevLett.66.994}
  {\path{doi:10.1103/PhysRevLett.66.994}}.

\bibitem{Joshi:1993zg}
P.~S. Joshi and I.~H. Dwivedi.
\newblock {Naked singularities in spherically symmetric inhomogeneous
  Tolman-Bondi dust cloud collapse}.
\newblock {\em Phys. Rev. D}, 47:5357--5369, 1993.
\newblock \href {http://arxiv.org/abs/gr-qc/9303037}
  {\path{arXiv:gr-qc/9303037}}, \href
  {https://doi.org/10.1103/PhysRevD.47.5357}
  {\path{doi:10.1103/PhysRevD.47.5357}}.

\bibitem{Harada:1998cq}
Tomohiro Harada, Hideo Iguchi, and Ken-ichi Nakao.
\newblock {Naked singularity formation in the collapse of a spherical cloud of
  counter rotating particles}.
\newblock {\em Phys. Rev. D}, 58:041502, 1998.
\newblock \href {http://arxiv.org/abs/gr-qc/9805071}
  {\path{arXiv:gr-qc/9805071}}, \href
  {https://doi.org/10.1103/PhysRevD.58.041502}
  {\path{doi:10.1103/PhysRevD.58.041502}}.

\bibitem{Joshi:2001xi}
Pankaj~S. Joshi, Naresh Dadhich, and Roy Maartens.
\newblock {Why do naked singularities form in gravitational collapse?}
\newblock {\em Phys. Rev. D}, 65:101501, 2002.
\newblock \href {http://arxiv.org/abs/gr-qc/0109051}
  {\path{arXiv:gr-qc/0109051}}, \href
  {https://doi.org/10.1103/PhysRevD.65.101501}
  {\path{doi:10.1103/PhysRevD.65.101501}}.

\bibitem{Goswami:2006ph}
Rituparno Goswami and Pankaj~S Joshi.
\newblock {Spherical gravitational collapse in N-dimensions}.
\newblock {\em Phys. Rev. D}, 76:084026, 2007.
\newblock \href {http://arxiv.org/abs/gr-qc/0608136}
  {\path{arXiv:gr-qc/0608136}}, \href
  {https://doi.org/10.1103/PhysRevD.76.084026}
  {\path{doi:10.1103/PhysRevD.76.084026}}.

\bibitem{Banerjee:2017njk}
Narayan Banerjee and Soumya Chakrabarti.
\newblock {Self-similar scalar field collapse}.
\newblock {\em Phys. Rev. D}, 95(2):024015, 2017.
\newblock \href {http://arxiv.org/abs/1701.04235} {\path{arXiv:1701.04235}},
  \href {https://doi.org/10.1103/PhysRevD.95.024015}
  {\path{doi:10.1103/PhysRevD.95.024015}}.

\bibitem{Bhattacharya:2017chr}
Kaushik Bhattacharya, Dipanjan Dey, Arindam Mazumdar, and Tapobrata Sarkar.
\newblock {New class of naked singularities and their observational
  signatures}.
\newblock {\em Phys. Rev. D}, 101(4):043005, 2020.
\newblock \href {http://arxiv.org/abs/1709.03798} {\path{arXiv:1709.03798}},
  \href {https://doi.org/10.1103/PhysRevD.101.043005}
  {\path{doi:10.1103/PhysRevD.101.043005}}.

\bibitem{Virbhadra:2002ju}
K.~S. Virbhadra and G.~F.~R. Ellis.
\newblock {Gravitational lensing by naked singularities}.
\newblock {\em Phys. Rev. D}, 65:103004, 2002.
\newblock \href {https://doi.org/10.1103/PhysRevD.65.103004}
  {\path{doi:10.1103/PhysRevD.65.103004}}.

\bibitem{Virbhadra:2007kw}
K.~S. Virbhadra and C.~R. Keeton.
\newblock {Time delay and magnification centroid due to gravitational lensing
  by black holes and naked singularities}.
\newblock {\em Phys. Rev. D}, 77:124014, 2008.
\newblock \href {http://arxiv.org/abs/0710.2333} {\path{arXiv:0710.2333}},
  \href {https://doi.org/10.1103/PhysRevD.77.124014}
  {\path{doi:10.1103/PhysRevD.77.124014}}.

\bibitem{Gyulchev:2008ff}
Galin~N. Gyulchev and Stoytcho~S. Yazadjiev.
\newblock {Gravitational Lensing by Rotating Naked Singularities}.
\newblock {\em Phys. Rev. D}, 78:083004, 2008.
\newblock \href {http://arxiv.org/abs/0806.3289} {\path{arXiv:0806.3289}},
  \href {https://doi.org/10.1103/PhysRevD.78.083004}
  {\path{doi:10.1103/PhysRevD.78.083004}}.

\bibitem{Sahu:2012er}
Satyabrata Sahu, Mandar Patil, D.~Narasimha, and Pankaj~S. Joshi.
\newblock {Can strong gravitational lensing distinguish naked singularities
  from black holes?}
\newblock {\em Phys. Rev. D}, 86:063010, 2012.
\newblock \href {http://arxiv.org/abs/1206.3077} {\path{arXiv:1206.3077}},
  \href {https://doi.org/10.1103/PhysRevD.86.063010}
  {\path{doi:10.1103/PhysRevD.86.063010}}.

\bibitem{Banerjee:2018clz}
Pritam Banerjee, Suvankar Paul, and Tapobrata Sarkar.
\newblock {On Strong Gravitational Lensing in Rotating Galactic Space-times}.
\newblock 4 2018.
\newblock \href {http://arxiv.org/abs/1804.07030} {\path{arXiv:1804.07030}}.

\bibitem{Shaikh:2019itn}
Rajibul Shaikh, Pritam Banerjee, Suvankar Paul, and Tapobrata Sarkar.
\newblock {Analytical approach to strong gravitational lensing from
  ultracompact objects}.
\newblock {\em Phys. Rev. D}, 99(10):104040, 2019.
\newblock \href {http://arxiv.org/abs/1903.08211} {\path{arXiv:1903.08211}},
  \href {https://doi.org/10.1103/PhysRevD.99.104040}
  {\path{doi:10.1103/PhysRevD.99.104040}}.

\bibitem{Paul:2020ufc}
Suvankar Paul.
\newblock {Strong gravitational lensing by a strongly naked null singularity}.
\newblock {\em Phys. Rev. D}, 102(6):064045, 2020.
\newblock \href {http://arxiv.org/abs/2007.05509} {\path{arXiv:2007.05509}},
  \href {https://doi.org/10.1103/PhysRevD.102.064045}
  {\path{doi:10.1103/PhysRevD.102.064045}}.

\bibitem{Tsukamoto:2021fsz}
Naoki Tsukamoto.
\newblock {Gravitational lensing by a photon sphere in a Reissner-Nordstr\"om
  naked singularity spacetime in strong deflection limits}.
\newblock {\em Phys. Rev. D}, 104(12):124016, 2021.
\newblock \href {http://arxiv.org/abs/2107.07146} {\path{arXiv:2107.07146}},
  \href {https://doi.org/10.1103/PhysRevD.104.124016}
  {\path{doi:10.1103/PhysRevD.104.124016}}.

\bibitem{Born:1934gh}
M.~Born and L.~Infeld.
\newblock {Foundations of the new field theory}.
\newblock {\em Proc. Roy. Soc. Lond. A}, 144(852):425--451, 1934.
\newblock \href {https://doi.org/10.1098/rspa.1934.0059}
  {\path{doi:10.1098/rspa.1934.0059}}.

\bibitem{Dey:2004yt}
Tanay~Kr. Dey.
\newblock {Born-Infeld black holes in the presence of a cosmological constant}.
\newblock {\em Phys. Lett. B}, 595(1-4):484--490, 2004.
\newblock \href {http://arxiv.org/abs/hep-th/0406169}
  {\path{arXiv:hep-th/0406169}}, \href
  {https://doi.org/10.1016/j.physletb.2004.06.047}
  {\path{doi:10.1016/j.physletb.2004.06.047}}.

\bibitem{Cai:2004eh}
Rong-Gen Cai, Da-Wei Pang, and Anzhong Wang.
\newblock {Born-Infeld black holes in (A)dS spaces}.
\newblock {\em Phys. Rev. D}, 70:124034, 2004.
\newblock \href {http://arxiv.org/abs/hep-th/0410158}
  {\path{arXiv:hep-th/0410158}}, \href
  {https://doi.org/10.1103/PhysRevD.70.124034}
  {\path{doi:10.1103/PhysRevD.70.124034}}.

\bibitem{Fernando:2003tz}
Sharmanthie Fernando and Don Krug.
\newblock {Charged black hole solutions in Einstein-Born-Infeld gravity with a
  cosmological constant}.
\newblock {\em Gen. Rel. Grav.}, 35:129--137, 2003.
\newblock \href {http://arxiv.org/abs/hep-th/0306120}
  {\path{arXiv:hep-th/0306120}}, \href
  {https://doi.org/10.1023/A:1021315214180}
  {\path{doi:10.1023/A:1021315214180}}.

\bibitem{Banerjee:2010da}
Rabin Banerjee, Sumit Ghosh, and Dibakar Roychowdhury.
\newblock {New type of phase transition in Reissner Nordström--AdS black hole
  and its thermodynamic geometry}.
\newblock {\em Phys. Lett. B}, 696:156--162, 2011.
\newblock \href {http://arxiv.org/abs/1008.2644} {\path{arXiv:1008.2644}},
  \href {https://doi.org/10.1016/j.physletb.2010.12.010}
  {\path{doi:10.1016/j.physletb.2010.12.010}}.

\bibitem{Zou:2013owa}
De-Cheng Zou, Shao-Jun Zhang, and Bin Wang.
\newblock {Critical behavior of Born-Infeld AdS black holes in the extended
  phase space thermodynamics}.
\newblock {\em Phys. Rev. D}, 89(4):044002, Feb 2014.
\newblock URL: \url{https://link.aps.org/doi/10.1103/PhysRevD.89.044002}, \href
  {http://arxiv.org/abs/1311.7299} {\path{arXiv:1311.7299}}, \href
  {https://doi.org/10.1103/PhysRevD.89.044002}
  {\path{doi:10.1103/PhysRevD.89.044002}}.

\bibitem{Hendi:2015hoa}
Seyed~Hossein Hendi, Behzad Eslam~Panah, and Shahram Panahiyan.
\newblock {Einstein-Born-Infeld-Massive Gravity: adS-Black Hole Solutions and
  their Thermodynamical properties}.
\newblock {\em JHEP}, 11:157, 2015.
\newblock \href {http://arxiv.org/abs/1508.01311} {\path{arXiv:1508.01311}},
  \href {https://doi.org/10.1007/JHEP11(2015)157}
  {\path{doi:10.1007/JHEP11(2015)157}}.

\bibitem{Zeng:2016sei}
Xiao-Xiong Zeng, Xian-Ming Liu, and Li-Fang Li.
\newblock {Phase structure of the Born--Infeld--anti-de Sitter black holes
  probed by non-local observables}.
\newblock {\em Eur. Phys. J. C}, 76(11):616, 2016.
\newblock \href {http://arxiv.org/abs/1601.01160} {\path{arXiv:1601.01160}},
  \href {https://doi.org/10.1140/epjc/s10052-016-4463-4}
  {\path{doi:10.1140/epjc/s10052-016-4463-4}}.

\bibitem{Li:2016nll}
Shoulong Li, H.~Lu, and Hao Wei.
\newblock {Dyonic (A)dS Black Holes in Einstein-Born-Infeld Theory in Diverse
  Dimensions}.
\newblock {\em JHEP}, 07:004, 2016.
\newblock \href {http://arxiv.org/abs/1606.02733} {\path{arXiv:1606.02733}},
  \href {https://doi.org/10.1007/JHEP07(2016)004}
  {\path{doi:10.1007/JHEP07(2016)004}}.

\bibitem{Tao:2017fsy}
Jun Tao, Peng Wang, and Haitang Yang.
\newblock {Testing holographic conjectures of complexity with Born--Infeld
  black holes}.
\newblock {\em Eur. Phys. J. C}, 77(12):817, 2017.
\newblock \href {http://arxiv.org/abs/1703.06297} {\path{arXiv:1703.06297}},
  \href {https://doi.org/10.1140/epjc/s10052-017-5395-3}
  {\path{doi:10.1140/epjc/s10052-017-5395-3}}.

\bibitem{Dehyadegari:2017hvd}
Amin Dehyadegari and Ahmad Sheykhi.
\newblock {Reentrant phase transition of Born-Infeld-AdS black holes}.
\newblock {\em Phys. Rev. D}, 98(2):024011, 2018.
\newblock \href {http://arxiv.org/abs/1711.01151} {\path{arXiv:1711.01151}},
  \href {https://doi.org/10.1103/PhysRevD.98.024011}
  {\path{doi:10.1103/PhysRevD.98.024011}}.

\bibitem{Wang:2018xdz}
Peng Wang, Houwen Wu, and Haitang Yang.
\newblock {Thermodynamics and Phase Transitions of Nonlinear Electrodynamics
  Black Holes in an Extended Phase Space}.
\newblock {\em JCAP}, 04(04):052, 2019.
\newblock \href {http://arxiv.org/abs/1808.04506} {\path{arXiv:1808.04506}},
  \href {https://doi.org/10.1088/1475-7516/2019/04/052}
  {\path{doi:10.1088/1475-7516/2019/04/052}}.

\bibitem{Liang:2019dni}
Kangkai Liang, Peng Wang, Houwen Wu, and Mingtao Yang.
\newblock {Phase structures and transitions of Born--Infeld black holes in a
  grand canonical ensemble}.
\newblock {\em Eur. Phys. J. C}, 80(3):187, 2020.
\newblock \href {http://arxiv.org/abs/1907.00799} {\path{arXiv:1907.00799}},
  \href {https://doi.org/10.1140/epjc/s10052-020-7750-z}
  {\path{doi:10.1140/epjc/s10052-020-7750-z}}.

\bibitem{Gan:2019jac}
Qingyu Gan, Guangzhou Guo, Peng Wang, and Houwen Wu.
\newblock {Strong cosmic censorship for a scalar field in a
  Born-Infeld\textendash{}de Sitter black hole}.
\newblock {\em Phys. Rev. D}, 100(12):124009, 2019.
\newblock \href {http://arxiv.org/abs/1907.04466} {\path{arXiv:1907.04466}},
  \href {https://doi.org/10.1103/PhysRevD.100.124009}
  {\path{doi:10.1103/PhysRevD.100.124009}}.

\bibitem{Wang:2019kxp}
Peng Wang, Houwen Wu, and Haitang Yang.
\newblock {Thermodynamics and Phase Transition of a Nonlinear Electrodynamics
  Black Hole in a Cavity}.
\newblock {\em JHEP}, 07:002, 2019.
\newblock \href {http://arxiv.org/abs/1901.06216} {\path{arXiv:1901.06216}},
  \href {https://doi.org/10.1007/JHEP07(2019)002}
  {\path{doi:10.1007/JHEP07(2019)002}}.

\bibitem{Wang:2020ohb}
Peng Wang, Houwen Wu, and Haitang Yang.
\newblock {Scalarized Einstein-Born-Infeld black holes}.
\newblock {\em Phys. Rev. D}, 103(10):104012, 2021.
\newblock \href {http://arxiv.org/abs/2012.01066} {\path{arXiv:2012.01066}},
  \href {https://doi.org/10.1103/PhysRevD.103.104012}
  {\path{doi:10.1103/PhysRevD.103.104012}}.

\bibitem{Guo:2022ghl}
Guangzhou Guo, Yuhang Lu, Peng Wang, Houwen Wu, and Haitang Yang.
\newblock {Black Holes with Multiple Photon Spheres}.
\newblock 12 2022.
\newblock \href {http://arxiv.org/abs/2212.12901} {\path{arXiv:2212.12901}}.

\bibitem{Novello:1999pg}
M.~Novello, V.~A. De~Lorenci, J.~M. Salim, and Renato Klippert.
\newblock {Geometrical aspects of light propagation in nonlinear
  electrodynamics}.
\newblock {\em Phys. Rev. D}, 61:045001, 2000.
\newblock \href {http://arxiv.org/abs/gr-qc/9911085}
  {\path{arXiv:gr-qc/9911085}}, \href
  {https://doi.org/10.1103/PhysRevD.61.045001}
  {\path{doi:10.1103/PhysRevD.61.045001}}.

\bibitem{Gan:2021xdl}
Qingyu Gan, Peng Wang, Houwen Wu, and Haitang Yang.
\newblock {Photon ring and observational appearance of a hairy black hole}.
\newblock {\em Phys. Rev. D}, 104(4):044049, 2021.
\newblock \href {http://arxiv.org/abs/2105.11770} {\path{arXiv:2105.11770}},
  \href {https://doi.org/10.1103/PhysRevD.104.044049}
  {\path{doi:10.1103/PhysRevD.104.044049}}.

\bibitem{Gan:2021pwu}
Qingyu Gan, Peng Wang, Houwen Wu, and Haitang Yang.
\newblock {Photon spheres and spherical accretion image of a hairy black hole}.
\newblock {\em Phys. Rev. D}, 104(2):024003, 2021.
\newblock \href {http://arxiv.org/abs/2104.08703} {\path{arXiv:2104.08703}},
  \href {https://doi.org/10.1103/PhysRevD.104.024003}
  {\path{doi:10.1103/PhysRevD.104.024003}}.

\bibitem{Guo:2022muy}
Guangzhou Guo, Xin Jiang, Peng Wang, and Houwen Wu.
\newblock {Gravitational lensing by black holes with multiple photon spheres}.
\newblock {\em Phys. Rev. D}, 105(12):124064, 2022.
\newblock \href {http://arxiv.org/abs/2204.13948} {\path{arXiv:2204.13948}},
  \href {https://doi.org/10.1103/PhysRevD.105.124064}
  {\path{doi:10.1103/PhysRevD.105.124064}}.

\bibitem{Chen:2022qrw}
Yiqian Chen, Guangzhou Guo, Peng Wang, Houwen Wu, and Haitang Yang.
\newblock {Appearance of an infalling star in black holes with multiple photon
  spheres}.
\newblock {\em Sci. China Phys. Mech. Astron.}, 65(12):120412, 2022.
\newblock \href {http://arxiv.org/abs/2206.13705} {\path{arXiv:2206.13705}},
  \href {https://doi.org/10.1007/s11433-022-1986-x}
  {\path{doi:10.1007/s11433-022-1986-x}}.

\bibitem{Guo:2021enm}
Guangzhou Guo, Peng Wang, Houwen Wu, and Haitang Yang.
\newblock {Quasinormal modes of black holes with multiple photon spheres}.
\newblock {\em JHEP}, 06:060, 2022.
\newblock \href {http://arxiv.org/abs/2112.14133} {\path{arXiv:2112.14133}},
  \href {https://doi.org/10.1007/JHEP06(2022)060}
  {\path{doi:10.1007/JHEP06(2022)060}}.

\bibitem{Guo:2022umh}
Guangzhou Guo, Peng Wang, Houwen Wu, and Haitang Yang.
\newblock {Echoes from hairy black holes}.
\newblock {\em JHEP}, 06:073, 2022.
\newblock \href {http://arxiv.org/abs/2204.00982} {\path{arXiv:2204.00982}},
  \href {https://doi.org/10.1007/JHEP06(2022)073}
  {\path{doi:10.1007/JHEP06(2022)073}}.

\bibitem{Guo:2023ivz}
Guangzhou Guo, Peng Wang, Houwen Wu, and Haitang Yang.
\newblock {Superradiance Instabilities of Charged Black Holes in
  Einstein-Maxwell-scalar Theory}.
\newblock 1 2023.
\newblock \href {http://arxiv.org/abs/2301.06483} {\path{arXiv:2301.06483}}.

\bibitem{Pugliese:2010ps}
D.~Pugliese, H.~Quevedo, and R.~Ruffini.
\newblock {Circular motion of neutral test particles in Reissner-Nordstr\"om
  spacetime}.
\newblock {\em Phys. Rev. D}, 83:024021, 2011.
\newblock \href {http://arxiv.org/abs/1012.5411} {\path{arXiv:1012.5411}},
  \href {https://doi.org/10.1103/PhysRevD.83.024021}
  {\path{doi:10.1103/PhysRevD.83.024021}}.

\bibitem{Bozza:2002zj}
V.~Bozza.
\newblock {Gravitational lensing in the strong field limit}.
\newblock {\em Phys. Rev. D}, 66:103001, 2002.
\newblock \href {http://arxiv.org/abs/gr-qc/0208075}
  {\path{arXiv:gr-qc/0208075}}, \href
  {https://doi.org/10.1103/PhysRevD.66.103001}
  {\path{doi:10.1103/PhysRevD.66.103001}}.

\bibitem{Tsukamoto:2016jzh}
Naoki Tsukamoto.
\newblock {Deflection angle in the strong deflection limit in a general
  asymptotically flat, static, spherically symmetric spacetime}.
\newblock {\em Phys. Rev. D}, 95(6):064035, 2017.
\newblock \href {http://arxiv.org/abs/1612.08251} {\path{arXiv:1612.08251}},
  \href {https://doi.org/10.1103/PhysRevD.95.064035}
  {\path{doi:10.1103/PhysRevD.95.064035}}.

\bibitem{Wei:2014dka}
Shao-Wen Wei, Ke~Yang, and Yu-Xiao Liu.
\newblock {Black hole solution and strong gravitational lensing in
  Eddington-inspired Born\textendash{}Infeld gravity}.
\newblock {\em Eur. Phys. J. C}, 75:253, 2015.
\newblock [Erratum: Eur.Phys.J.C 75, 331 (2015)].
\newblock \href {http://arxiv.org/abs/1405.2178} {\path{arXiv:1405.2178}},
  \href {https://doi.org/10.1140/epjc/s10052-015-3556-9}
  {\path{doi:10.1140/epjc/s10052-015-3556-9}}.

\bibitem{Cunha:2016bpi}
Pedro V.~P. Cunha, Carlos A.~R. Herdeiro, Eugen Radu, and Helgi~F. Runarsson.
\newblock {Shadows of Kerr black holes with and without scalar hair}.
\newblock {\em Int. J. Mod. Phys. D}, 25(09):1641021, 2016.
\newblock \href {http://arxiv.org/abs/1605.08293} {\path{arXiv:1605.08293}},
  \href {https://doi.org/10.1142/S0218271816410212}
  {\path{doi:10.1142/S0218271816410212}}.

\bibitem{Liu:2019rib}
Hai-Shan Liu, Zhan-Feng Mai, Yue-Zhou Li, and H.~L\"u.
\newblock {Quasi-topological Electromagnetism: Dark Energy, Dyonic Black Holes,
  Stable Photon Spheres and Hidden Electromagnetic Duality}.
\newblock {\em Sci. China Phys. Mech. Astron.}, 63:240411, 2020.
\newblock \href {http://arxiv.org/abs/1907.10876} {\path{arXiv:1907.10876}},
  \href {https://doi.org/10.1007/s11433-019-1446-1}
  {\path{doi:10.1007/s11433-019-1446-1}}.

\bibitem{Johnson:2019ljv}
Michael~D. Johnson et~al.
\newblock {Universal interferometric signatures of a black
  hole\textquoteright{}s photon ring}.
\newblock {\em Sci. Adv.}, 6(12):eaaz1310, 2020.
\newblock \href {http://arxiv.org/abs/1907.04329} {\path{arXiv:1907.04329}},
  \href {https://doi.org/10.1126/sciadv.aaz1310}
  {\path{doi:10.1126/sciadv.aaz1310}}.

\bibitem{Himwich:2020msm}
Elizabeth Himwich, Michael~D. Johnson, Alexandru Lupsasca, and Andrew
  Strominger.
\newblock {Universal polarimetric signatures of the black hole photon ring}.
\newblock {\em Phys. Rev. D}, 101(8):084020, 2020.
\newblock \href {http://arxiv.org/abs/2001.08750} {\path{arXiv:2001.08750}},
  \href {https://doi.org/10.1103/PhysRevD.101.084020}
  {\path{doi:10.1103/PhysRevD.101.084020}}.

\bibitem{Gralla:2020srx}
Samuel~E. Gralla, Alexandru Lupsasca, and Daniel~P. Marrone.
\newblock {The shape of the black hole photon ring: A precise test of
  strong-field general relativity}.
\newblock {\em Phys. Rev. D}, 102(12):124004, 2020.
\newblock \href {http://arxiv.org/abs/2008.03879} {\path{arXiv:2008.03879}},
  \href {https://doi.org/10.1103/PhysRevD.102.124004}
  {\path{doi:10.1103/PhysRevD.102.124004}}.

\end{thebibliography}

\end{document}